\documentclass[%
 reprint,
preprintnumbers,
 amsmath,amssymb,
 aps,
]{revtex4-1}

\usepackage{graphicx}
\usepackage{dcolumn}
\usepackage{bm}
\usepackage{hyperref}
\usepackage{siunitx}
\usepackage{multirow}
\usepackage{subfig}
\usepackage[english]{babel}
\usepackage{booktabs}
\usepackage{algorithm}
\usepackage{algorithmic}

\newcommand*{\dif}{\mathop{}\!\mathrm{d}}

\begin{document}


\title{Longitudinal Localized Kick Driven Fast Extraction Method and Rapid Cycling Synchrotron Design for 3D PBS Proton FLASH Delivery}

\author{Yang Xiong}
\author{Hongjuan Yao}
\author{Shuxin Zheng}
\email{Contact author: zhengsx@mail.tsinghua.edu.cn}
\affiliation{%
 Department of Engineering Physics, Tsinghua University, Beijing 100084, China\\
 Key Laboratory of Particle $\&$ Radiation Imaging (Tsinghua University), Ministry of Education, Beijing 100084, China\\
 Laboratory for Advanced Radiation Source and Application, Tsinghua University, Beijing 100084, China
}%

%
%


\begin{abstract}
This paper presents the design of a rapid cycling synchrotron (RCS) featuring a longitudinal localized kick driven fast extraction system for three-dimensional (3D) pencil beam scanning (PBS) proton FLASH delivery.
The extraction method is designed to accommodate a novel scanning scheme that addresses the stringent requirement for substantially shorter delivery time compared to current solutions, where the scanning layer is parallel to the proton beam direction.
In this method, the kicker pulse waveform is applied selectively to specific longitudinal segments of the proton bunch.
For each scanning spot, the functional region of the kicker along the longitudinal direction is dynamically adjusted based on real-time beam longitudinal line density measured by a beam current monitor.
The corresponding region-determination algorithm is provided.
We analyze the spot dose accuracy and the beam loss at the septum, indentifying increased particle longitudinal line density will reduce spot dose accuracy and increase beam loss.
A total number of particles of $2\times10^{10}$ can satisfy the requirements of spot dose accuracy and the beam loss due to the septum is less than 1\%.
The extraction system comprises a stripline kicker, an electric septum (ESe), and a magnetic septum (MSe), imposing specific requirements on the RCS lattice design.
The RCS is carefully designed to meet these constraints, and the parameters of the extraction elements are detailed.
By integrating a novel scanning scheme with a specially designed RCS and fast extraction method, this work demonstrates the feasibility of achieving 3D PBS proton FLASH delivery.

\end{abstract}

\maketitle


\section{Introduction}\label{Sec.1}
Ultra-high dose rate radiotherapy (FLASH-RT) has shown considerable promise in radiation oncology owing to its ability to reduce normal tissue toxicity while maintaining tumor control\cite{Favaudon_2014,Esplen_2020,Loo_2024}.
This technique requires delivering a sufficient dose rate ($>$ \SI{40}{Gy/s}) within an irradiation time on the order of \SI{100}{ms}.
For proton FLASH-RT, several beam delivery approaches can be employed.
Along the depth direction, either the plateau region\cite{10.1001/jamaoncol.2022.5843} or the Bragg peak\cite{Darafsheh_2025} of the dose deposition curve can be utilized.
While in the transverse direction, both beam broadening with one or two scatterers\cite{DIFFENDERFER2020440} and pencil beam scanning (PBS)\cite{Klimpki_2018} are viable options.
Among these, PBS utilizing the proton Bragg peak is regarded as the most promising delivery method, as it allows for intensity modulation and takes advantage of the high Linear Energy Transfer (LET) region of the Bragg peak.

Proton beam energy modulation is needed to form spread-out Bragg peak (SOBP) in depth direction.
In common scanning irradiation practice, the beam is perpendicular to the scanning layers.
After one layer is scanned, the beam energy is changed and the next layer is scanned.
When deeper layers are irradiated, the shallower layers also receive dose.
Consequently, for the shallowest layer, the total irradiation time equals the time required to irradiate the entire target volume.
So the energy switching time is a key concern.
For cyclotron-based systems, energy switching is achieved by adjusting the thickness of the energy degrader, a process that takes at least \SI{50}{ms}\cite{https://doi.org/10.1002/mp.13972}.
In medical synchrotrons, energy switching between cycles requires several seconds, while energy changing within a cycle still takes time on the order of hundreds of milliseconds\cite{YOUNKIN2018412}.
This makes the overall irradiation time far exceed \SI{100}{ms} for large targets requiring tens of scanning layers and unable to meet the FLASH dose rate criterion.

One solution is to use ridge filters, which enables SOBP generation within the target area despite using a monoenergetic beam\cite{Roberfroid_2025,RODDY2024169284}.
So transverse PBS combined with ridge filter is currently a widely used approach in proton FLASH study.
However, ridge filter systems need to be specially customized for each patient.
Even though for the same patient, new ridge filter is also needed with adaptation to daily anatomy.
Besides, proton beam passing through ridge filters may cause additional neutron doses for the patients.

\begin{figure}[!htb]
	\centering
	\includegraphics*[width=0.99\columnwidth]{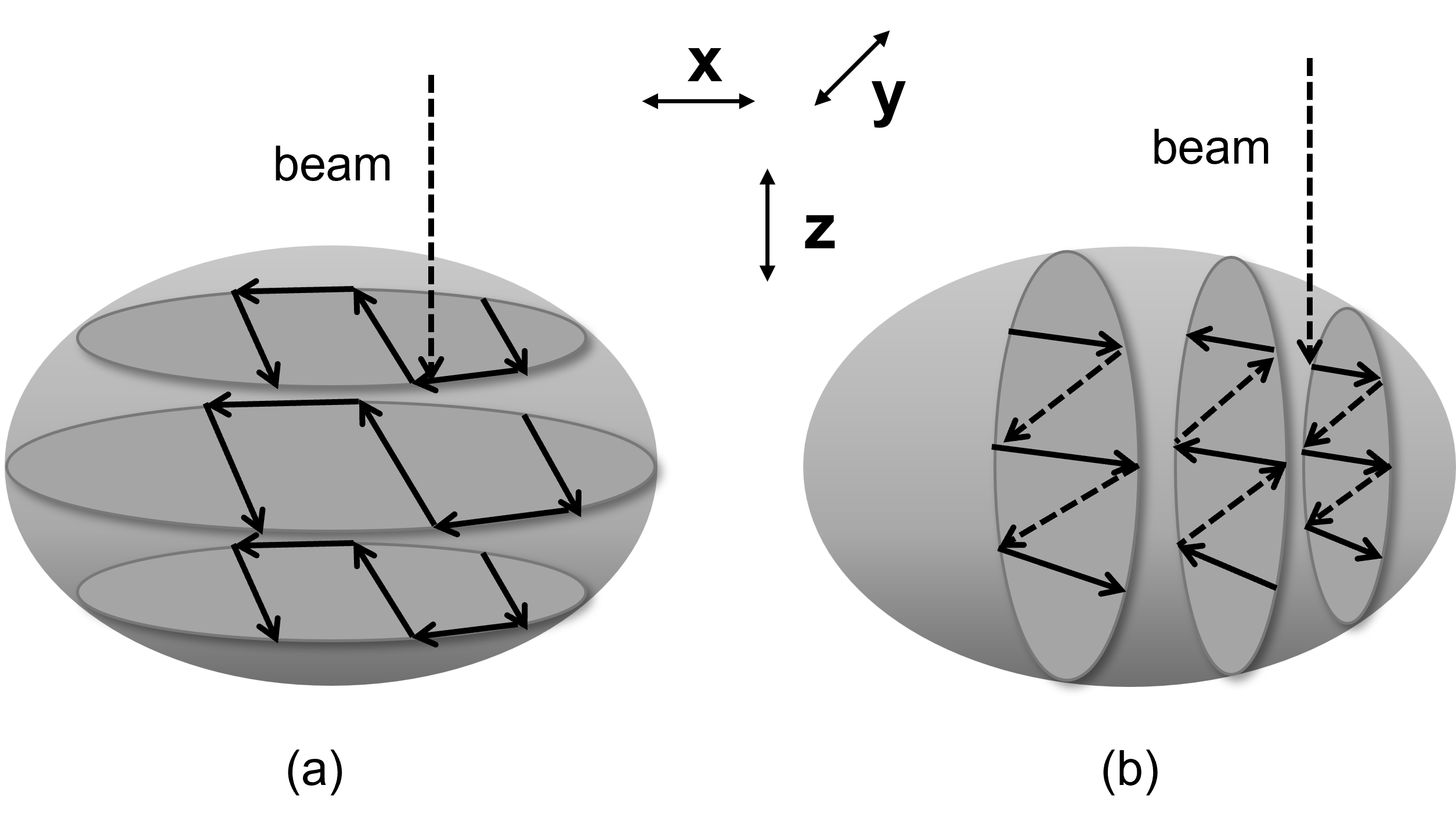}
	\caption{Two choices for the beam delivery approach\cite{Xiong_2025}. (a) scanning layer is oriented perpendicular to the beam direction. After a layer is scanned, the beam energy is changed and the next layer is being scanned. (b) scanning layer is parallel to the beam direction. The beam is scanned in both transverse direction and depth direction. When the scanning of a layer is complete, the beam moves transversely and the next adjacent layer will be scanned.}
	\label{fig:Layering_Scheme}
\end{figure}

Another approach is to align the beam parallel to the scanning layer\cite{JayFlanz2022,Xiong_2025}, the beam is scanned both transversely within the layer and the beam energy is changed to finish the depth direction scanning, as shown in Fig.~\ref{fig:Layering_Scheme}b.
In this new scanning pattern where the beam is positioned within a layer, when a certain layer is being irradiated, the distant layers remain unexposed.
Here we use the concept of ‘effective irradiation time’\cite{https://doi.org/10.1002/mp.14456} to describe the dosimetry of voxels inside the tumour target volume.
Generally the cross‑section of the beam can cover around 5 adjacent scanning layers with \SI{5}{mm} layer thickness\cite{https://doi.org/10.1002/acm2.12984}, then effective irradiation time per voxel equals the time needed to scan those 5 layers.
As a contrast, in the schematic of proton beam perpendicular to the scanning layer, the effective irradiation time equals the time needed to scan all layers of the target.

\begin{figure*}[!htb]
	\centering
	\includegraphics*[width=1.99\columnwidth]{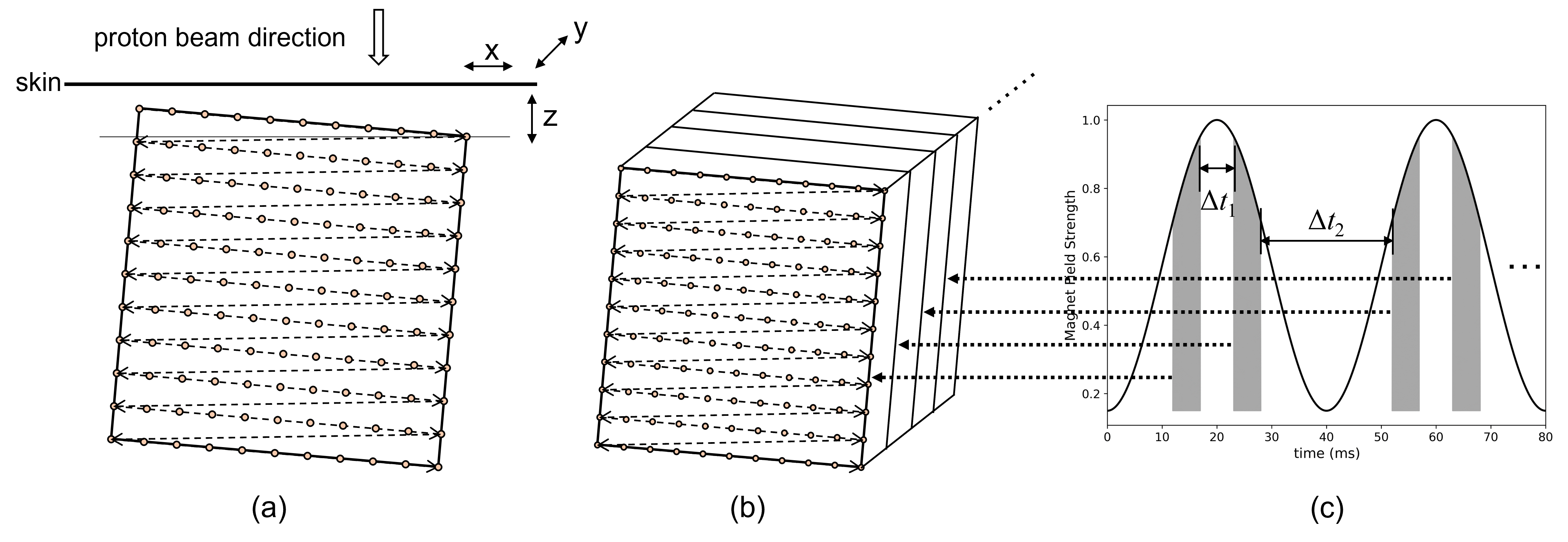}
	\caption{Scanning pattern schematic\cite{Xiong_2025}. (a) 2D scanning pattern within a single layer ($x$–$z$ plane). Motion in the $z$‑direction is achieved by varying the beam energy; motion in the $x$‑direction is controlled by the scanning magnet. (b) 3D illustration of the multi‑layer coverage. (c) Sine‑wave of the bending‑magnet field during one RCS cycle.}
	\label{fig:Scanning_Pattern}
\end{figure*}

Then the aim is to finish the scanning of 5 layers within around \SI{100}{ms}.
Rapid cycling synchrotron (RCS) provides a promising solution, the reasons are as follows:
First, the scanning in depth direction within a layer requires the accelerator to provide proton beams with variable energy, only synchrotrons can address this;
second, calculation shows that to scan a \SI{1.0}{L} target volume with a total dose of \SI{8}{Gy}, particles needed is around 10 times larger than a single synchrotron cycle storage limitation.
Therefore, common slow cycling synchrotrons are not usable due to their long cycle gap.
RCS with substantially shorter cycle time works.
For an RCS operating at \SI{25}{Hz}, the duration of one cycle is only \SI{40}{ms}\cite{CHEN2025170431}, and Fig.~\ref{fig:Scanning_Pattern}c shows the sine wave of the magnetic field of the bending magnets in the RCS ring.

The scanning pattern of this novel scheme within a layer and between layers are shown in Fig.~\ref{fig:Scanning_Pattern}a and b.
The proton beam extracted at different times of the RCS cycle possesses different energies, corresponding to different depths in tissue.
So in Fig.~\ref{fig:Scanning_Pattern}a, the scanning in the $z$‑direction is achieved by extract the beam continuously during the RCS cycle and scanning in the $x$‑direction is realized by the scanning magnet.
In Fig.~\ref{fig:Scanning_Pattern}b, the switch between layers is also realized by scanning magnet deflecting the beam in another direction.
If the beams are extracted during the acceleration phase (rising part of the sine wave in Fig.~\ref{fig:Scanning_Pattern}c), the shallow part of the layer is irradiated first.
Conversely, if extraction occurs during the deceleration phase (falling part), the scanning sequence is reversed, beginning with the deeper part.
Thus, two layers can be scanned in a single RCS cycle (\SI{40}{ms}), enabling 5 scanning layers to be completed in only \SI{100}{ms}—thereby meeting the FLASH irradiation time requirement.
A detailed description and supporting calculations of this scheme can be found in Ref.~\cite{Xiong_2025}.

Typically, there is only one or two bunches are stored in the synchrotron ring.
So a key challenge of this scheme is to extract hundreds of individual beamlets from one bunch while maintaining spot dose accuracy within clinically acceptable limits.
The main content of this paper is to introduce a longitudinal localized kick driven fast extraction method to realize this.
The conceptual framework of this method is introduced, and particle tracking simulation results—focusing on spot dose accuracy and beam loss—are presented and discussed.

Currently, all of the medical hadron synchrotrons for cancer therapy operate in a slow cycling mode, with typical cycle time on the order of several seconds\cite{https://doi.org/10.1118/1.3187229,COMBS201041}.
No RCS has yet been deployed for clinical use, with only several designs proposed in earlier literature, including schemes from Brookhaven National Laboratory (BNL)\cite{Cardona1221935} and Toshiba Corporation\cite{Yamaguchi753429}.
The parameters of these two RCS designs are summarized in Table~\ref{tab:RCS_design}.
\begin{table}[!htb]
	\caption{\label{tab:RCS_design}
		Design parameters of the RCS
	}
	\begin{ruledtabular}
		\begin{tabular}{ccc}
			& BNL & Toshiba\\
			\colrule
			Circumference             & \SI{28.6}{m}      & \SI{28.2}{m}     \\
			Repetation rate           & 60 or \SI{30}{Hz} & \SI{20}{Hz}      \\
			Maximum bending field     & -                 & \SI{1.2}{T}      \\
		\end{tabular}
	\end{ruledtabular}
\end{table}
However, the longitudinal localized kick driven fast extraction method to be discussed in this paper imposes specific design requirements on the RCS, particularly regarding the phase advance between key extraction elements such as the fast kicker and the electric and magnetic septa.
Consequently, this paper also presents the design of an RCS specifically tailored to this extraction method.

The remainder of this paper is structured as follows.
Following the introduction in Sec.~\ref{Sec.1}, Sec.~\ref{Sec.2} provides a detailed description of the longitudinal localized kick driven fast extraction method.
The design of the dedicated RCS is presented in Sec.~\ref{Sec.3}.
Sec.~\ref{Sec.4} details simulations performed based on this new RCS design, which validate the method's effectiveness, analyze the impact of longitudinal line density on spot dose accuracy, and give the beam loss associated with the electric septum.
A discussion is presented in Sec.~\ref{Sec.5} and a conclusion is given in Sec.~\ref{Sec.6}.

\section{Longitudinal Localized Kick Driven Fast Extraction method}\label{Sec.2}
\begin{figure}[!htb]
	\centering
	\includegraphics*[width=0.99\columnwidth]{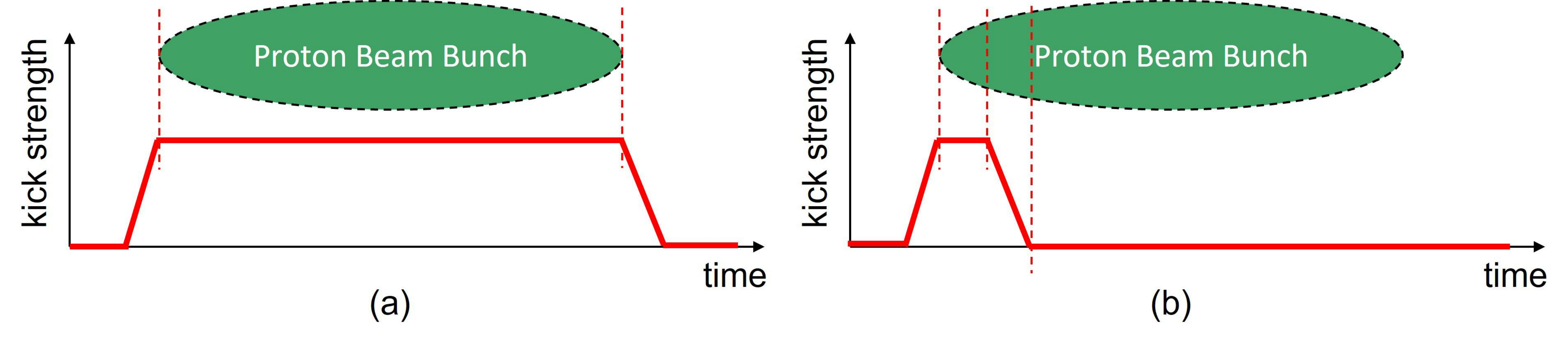}
	\caption{The relationship between the kicker pulse time and bunch length. (a) Conventional fast extraction, the entire bunch is kicked out in a single turn; (b) Fast extraction used in our scheme, the bunch needs to support hundreds of scanning spots.}
	\label{fig:Fast_Kick_Compare}
\end{figure}

Typical fast extraction system consists of a pulsed kicker and a septum.
The pulsed kicker deflects the beam over the septum blade and into the high-field region of the septum magnet (MSe) located at a suitable phase advance downstream\cite{Fraser:2018bcq}.
In the usual fast extraction system, the flattop of the kick pulse is applied to the entire bunch, with rising and falling edges locates at the longitudinal segments where no particles exists.
However, the novel scanning scheme needs to confine the kick to a specific longitudinal segment of the bunch to extract only a fraction of the bunch to finish the scanning of hundreds of spots within one layer.
The schematic of the fast kicker pulse difference is conceptually illustrated in Fig.~\ref{fig:Fast_Kick_Compare}.

Choosing the right type of fast kicker is also crucial.
Compact medical synchrotrons typically store bunches with lengths of several hundred nanoseconds, necessitating kicker rise/fall times on the order of nanoseconds.
Consequently, the stripline kicker was selected.
A stripline kicker, also called an electromagnetic kicker, consists of two parallel electrodes (striplines).
When a charged particle $q$ with energy $E$ traverses stripline electrodes of length $L$, it experiences a deflection angle,
\begin{equation}\label{eq:stripline_kicker_1}
	\Delta\theta=2g_{\perp}\left(\frac{qV}{E}\right)\frac{L}{d}
\end{equation}
relative to its reference or initial orbit. Here, $g_{\perp}$ is the transverse geometry factor given by
\begin{equation}\label{eq:stripline_kicker_2}
	g_{\perp}=\tanh\left(\frac{\pi w}{2d}\right)
\end{equation}
$V$ is the pulse voltage amplitude, $d$ is the gap between the parallel electrodes, and $w$ is the width of the stripline\cite{PhysRevAccelBeams.28.020401}.

Under current technology, the pulse voltage amplitude $V$ of the stripline is limited to around \SI{20}{kV}.
But the proton beam has a larger profile in the $x$-direction, resulting in a larger separation distance (correspond to larger deflection angle $\Delta\theta$) and gap between the parallel electrodes $d$.
Therefore, according to Eqs.~\eqref{eq:stripline_kicker_1} and \eqref{eq:stripline_kicker_2}, it is considered to apply the kick to the beam in the $y$-direction.

\begin{figure}[!htb]
	\centering
	\includegraphics*[width=0.99\columnwidth]{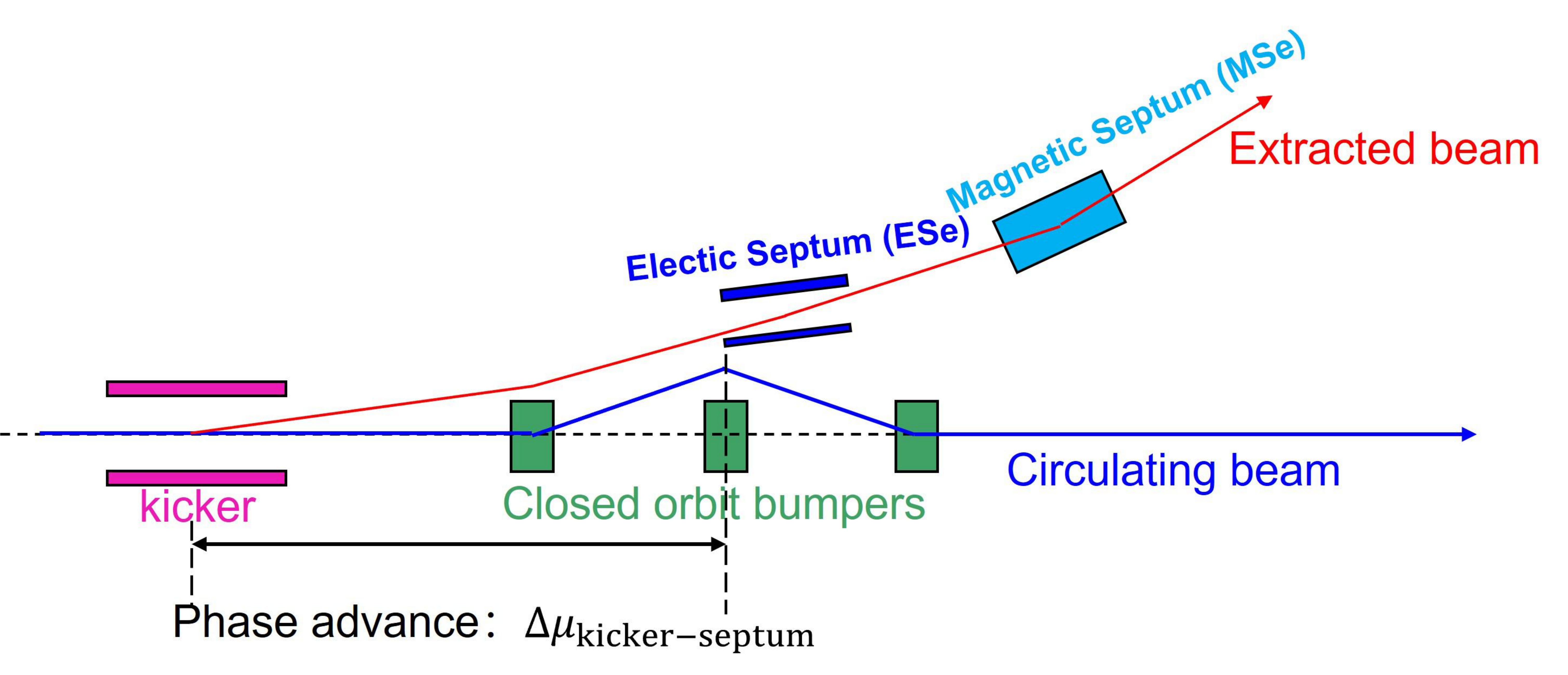}
	\caption{Schematic of fast extraction system and beam motion in the transverse normalized phase space.}
	\label{fig:Fast_Extraction_Layout}
\end{figure}

Generally the MSe has a septum wall with more than \SI{10}{mm} thickness\cite{WU201845}.
To further reduce the parameter requirements of the stripline kicker, an additional electric septum (ESe) with only \SI{0.1}{mm} septum wall thickness is added between the pulsed kicker and the final septum.
The overall layout of the fast extraction system is shown in Fig.~\ref{fig:Fast_Extraction_Layout}.
Particles are first kicked by the stripline kicker and move into the ESe, then being deflected by the ESe and move into the MSe, and finally enter the transport line.

For detailed calculation of the fast extraction system and element parameters, please refer to Sec.~\ref{Sec.3.B} and Sec.~\ref{Sec.3.C}.

\subsection{Particle Motion Analysis}
Precise spot dose control is of great significance in PBS proton therapy and must be carefully considered in the design.
Dose control at each scanning spot can be directly translated into controlling the number of particles extracted.
And this extracted particle count varies with the longitudinal segment where the kick pulse is applied.
Therefore, precise particle number control can be achieved by regulating the kicker's pulse time.

Accordingly, we employ a feedback method based on the particle line density along the longitudinal direction $\lambda\left(\tau\right)$ with unit of particles$\cdot \mathrm{m}^{-1}$ ($\tau$ is the time of arrival of nonsynchronous particle relative to the synchronous particle).
This approach involves measuring the proton beam current $I\left(t\right) = I\left(\tau + nT_{\mathrm{rev}}\right)$ ($n$ is the turn number and $T_{\mathrm{rev}}$ is the revolution period of synchronous particle) over time along the ring circumference using beam current monitor (BCM)\cite{Brandt:1071486}.
The relationship between longitudinal particle line density $\lambda\left(\tau\right)$ and current $I\left(\tau\right)$ in one revolution period can be written as:
\begin{equation}\label{I_vs._lambda}
	I\left(\tau\right) = e\lambda\left(\tau\right) \cdot \frac{\mathrm{Circ}}{T_{\mathrm{rev}}}
\end{equation}
where $e$ is the charge of a proton and $\mathrm{Circ}$ is the circumference of the synchrotron ring.
In the following part of this paper, we use $\lambda\left(\tau\right)$ to express the beam longitudinal distribution.

The rising edge can be aligned with a vacant longitudinal segment and no particles will be extracted, as shown in Fig.~\ref{fig:Fast_Kick_Compare}(b).
The number of extracted particles in the flattop duration and the fall time can be analyzed separately and we start from flattop duration.

\begin{figure*}[!htb]
	\centering
	\subfloat[\label{fig:TOP_ca_kicker_1_in_85}]
	{\includegraphics*[width=0.30\linewidth]{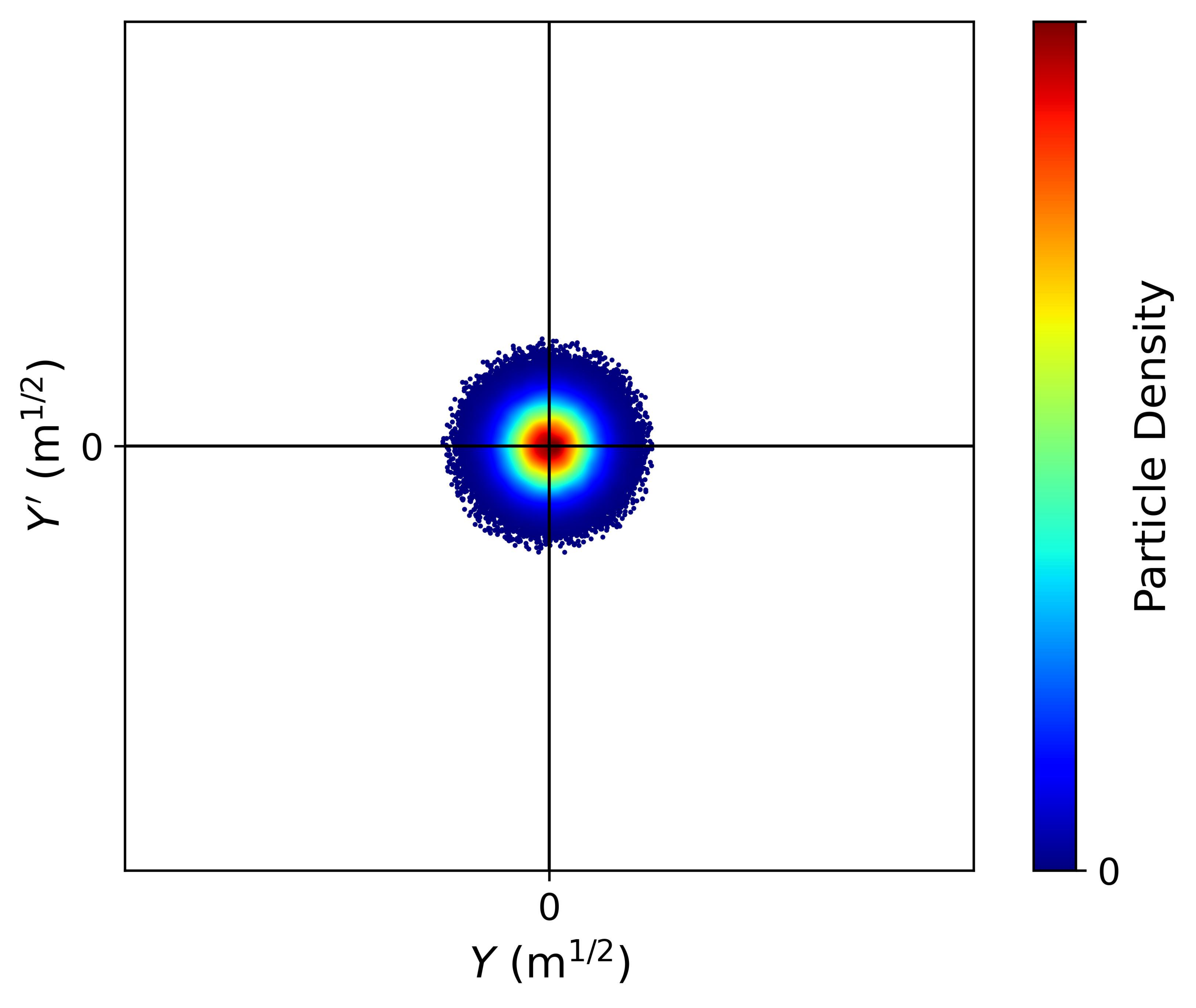}}
	\subfloat[\label{fig:TOP_ca_kicker_1_out_85}]
	{\includegraphics*[width=0.30\linewidth]{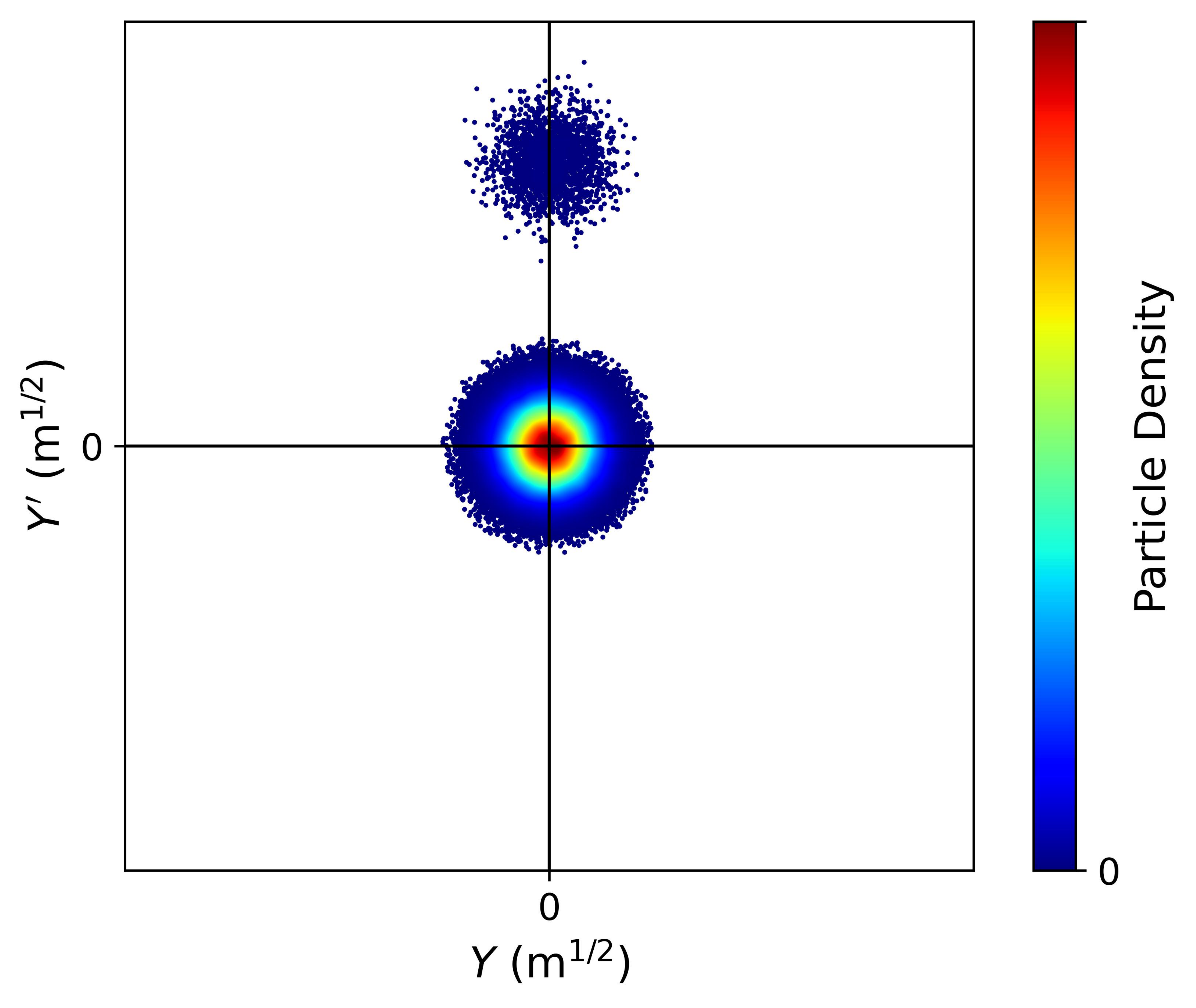}}
	\subfloat[\label{fig:TOP_ca_ese_in_85}]
	{\includegraphics*[width=0.30\linewidth]{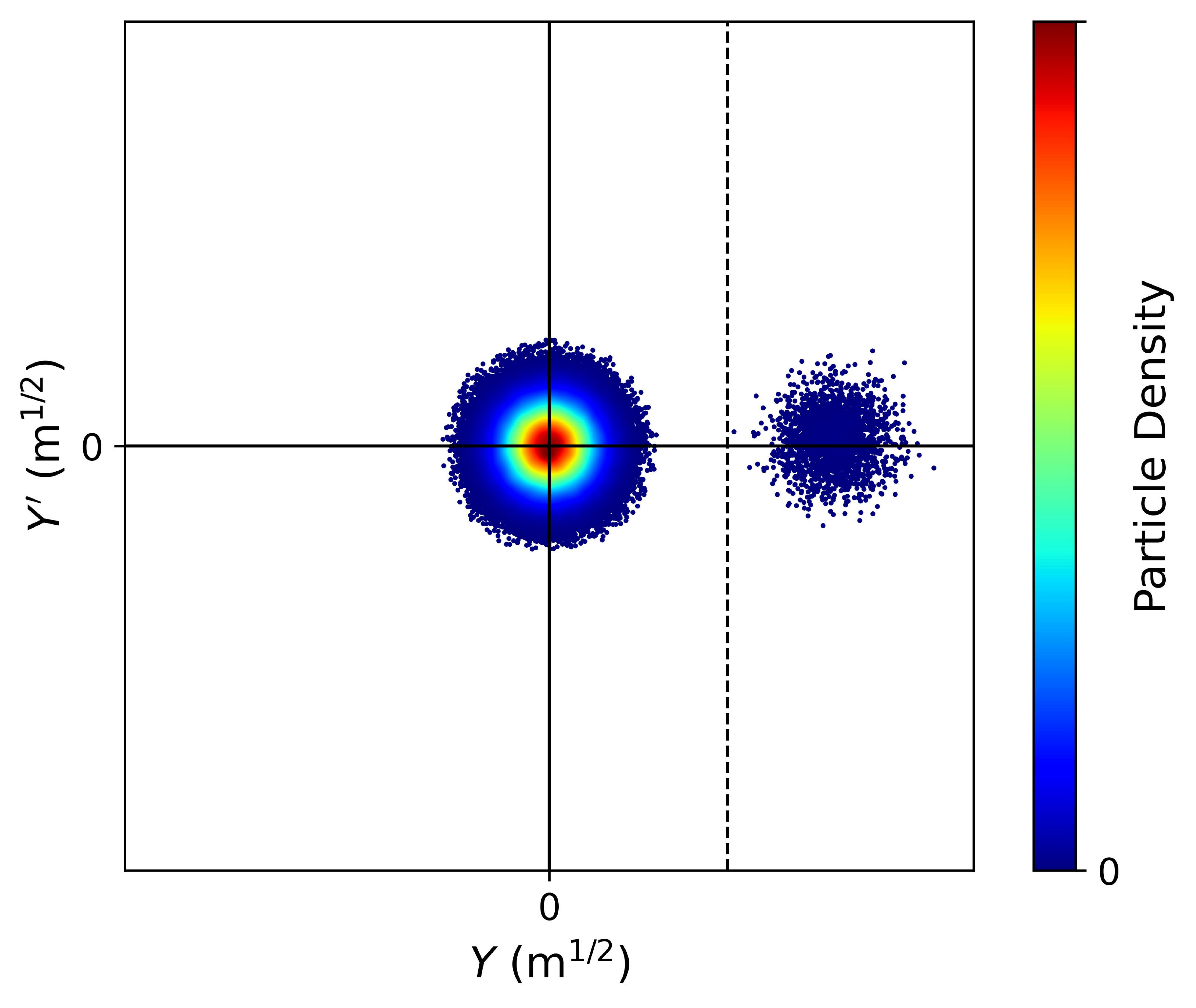}}
	\caption{Motion of particles extracted in flattop duration in normalized phase space. (a) Initial particle distribution (befored kicked by the stripline kicker). (b) particles in the falttop duration are kicked. (c) kicked particles move into the ESe (the vertical dashed line represents the $y$ coordinate of the ESe).}
	\label{fig:TOP}
\end{figure*}

The start time ($\tau_1 = n_1 \cdot \Delta\tau$) of the fast kicker pulse flattop is determined by Eq.~\eqref{eq:Flattop_Begin}:
\begin{equation}\label{eq:Flattop_Begin}
	\begin{aligned}
		\lambda\left( i\cdot\Delta\tau \right) &= 0, \forall i\in \mathbb{N}, i\le n_1\\
		\lambda\left( \left( n_1+1 \right) \cdot\Delta\tau \right) &> 0
	\end{aligned}
\end{equation}
and with $\tau_2 = n_2 \cdot \Delta\tau$ represents the end time of the pulse flattop, the particle number extracted in the flattop duration $N_{\mathrm{flattop}}$ can be expressed as
\begin{equation}\label{eq:Flattop_num}
	N_{\mathrm{flattop}} = \sum_{i=n_1+1}^{n_2} \lambda\left(i\cdot\Delta\tau\right) \cdot \Delta\tau
\end{equation}
The motion of particles extracted in the flattop duration in the normalized phase space ($Y-Y^{\prime}$) is illustrated in Fig.~\ref{fig:TOP}.


Then we consider the particles extracted during the pulse fall time.
The falling edge exhibits distinct behavior due to the time-dependent kick field.
Unlike the flattop, where particles receive a full kick, those distributed in the falling edge are not fully kicked and thus only partially extracted.

The analysis is also conducted in the $Y-Y^{\prime}$ phase space.
The separation distance of the bunch fully kicked during the flattop at the entrance of the ESe is expressed as $Y_{c}$ in the $Y-Y^{\prime}$ phase space.
The falling edge is assumed as linear and $\tau_2$ and $\tau_3$ denote the start and end time points of the falling edge respectively.
For $\tau_2<\tau<\tau_3$, the separation distance is then written as
\begin{equation}\label{eq:FALL_separation distance}
	Y_{c,\mathrm{fall}} = \frac{\tau_3-\tau}{\tau_3-\tau_2}Y_c
\end{equation}

To calculate the number of particles that enters the ESe, marginal distribution of the kicked bunch in the $Y$ direction at time $\tau$ during the fall time  $f_{Y}\left(Y,\tau\right)$ is needed.
Assuming a Gaussian transverse particle distribution, $f_{Y}\left(Y,\tau\right)$ can be written as:
\begin{equation}\label{eq:Y_distribution}
	f_{Y}\left(Y,\tau\right)  = \frac{1}{\sqrt{2\pi\varepsilon_{\mathrm{rms}}}} \mathrm{exp}\left[ -\frac{\left( Y-Y_{c,\mathrm{fall}} \right) ^2}{2\varepsilon_{\mathrm{rms}}} \right]
\end{equation}
where $\varepsilon_{\mathrm{rms}}$ is the rms transverse emittance of the extracted bunch.

Then for the falling edge with longitudinal particle line density of $\lambda\left(\tau\right)$, the number of particles extracted theoretically is 
\begin{equation}\label{eq:N_fall_expression}
	\begin{aligned}
		N_{\mathrm{fall}} &= \int_{\tau_2}^{\tau_3} \lambda\left(\tau\right) \cdot \frac{\mathrm{Circ}}{T_{\mathrm{rev}}} \dif \tau \int_{Y_{\mathrm{ESe}}}^{+\infty} f_{Y}\left(Y,\tau\right) \dif Y \\
		&= \frac{1}{2}\frac{\mathrm{Circ}}{T_{\mathrm{rev}}} \int_{\tau_2}^{\tau_3} \lambda\left(\tau\right) \cdot \mathrm{erfc}\left( \frac{Y_{\mathrm{ESe}}-\frac{\tau_3-\tau}{\tau_3-\tau_2}Y_{c,\mathrm{fall}}}{\sqrt{2\varepsilon_{\mathrm{rms}}}} \right) \dif \tau
	\end{aligned}
\end{equation}
where $Y_{\mathrm{ESe}}$ is the transverse location of the ESe anode at the entrance in the $Y-Y^{\prime}$ phase space.

\begin{figure*}[!htb]
	\centering
	\subfloat[\label{fig:TOP_FALL_ca_kicker_1_in_85}]
	{\includegraphics*[width=0.30\linewidth]{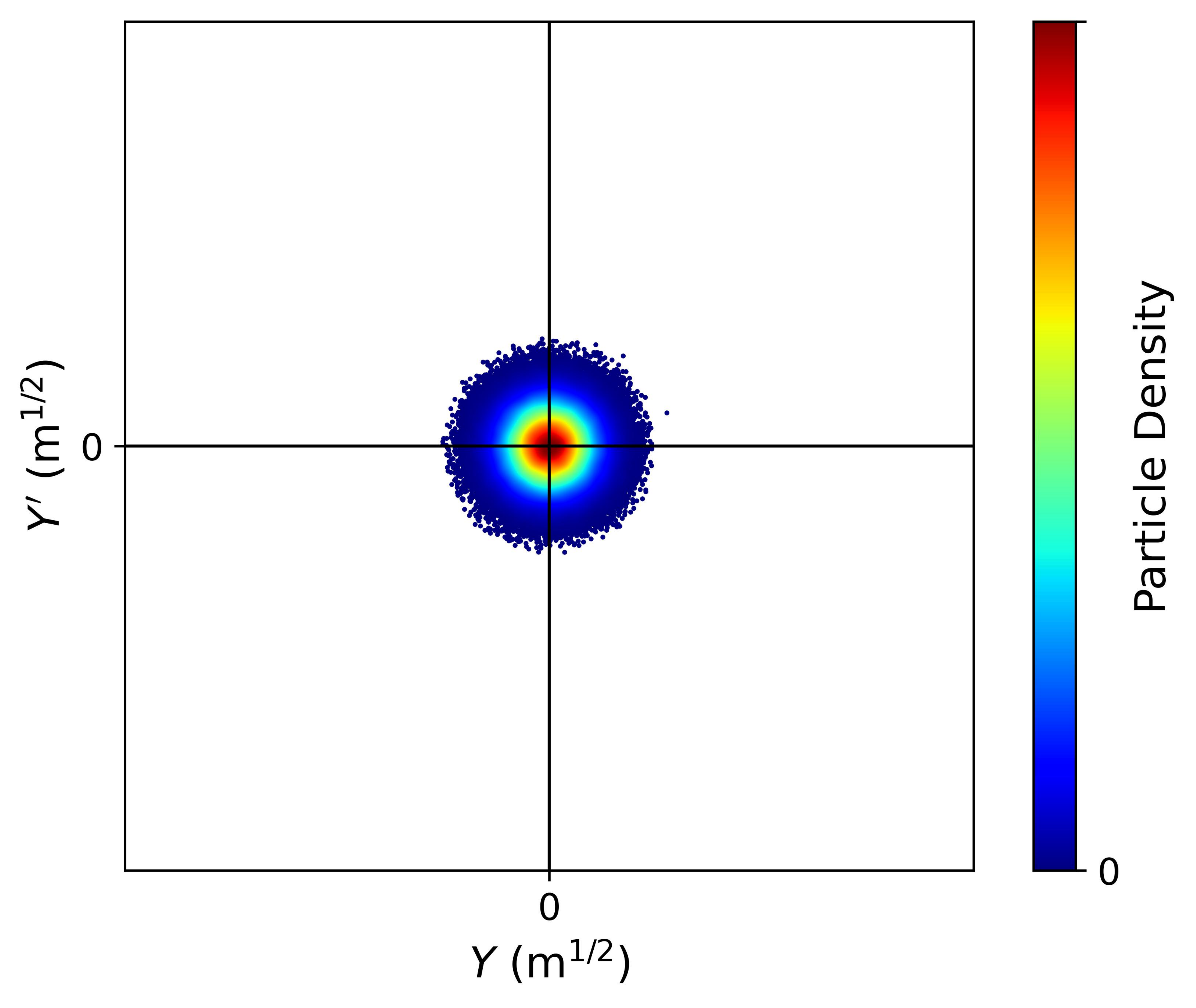}}
	\subfloat[\label{fig:TOP_FALL_ca_kicker_1_out_85}]
	{\includegraphics*[width=0.30\linewidth]{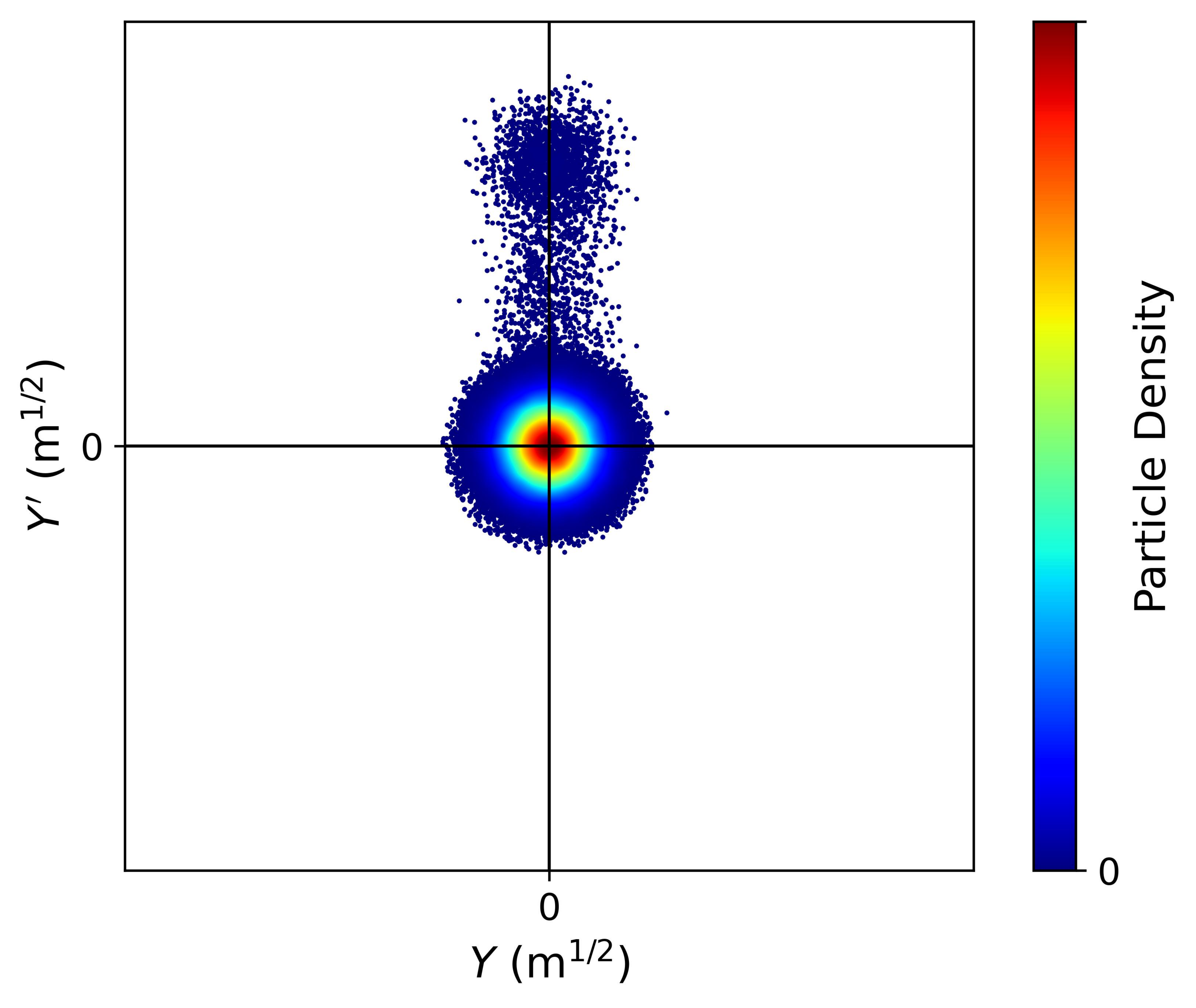}}
	\subfloat[\label{fig:TOP_FALL_ca_ese_in_85}]
	{\includegraphics*[width=0.30\linewidth]{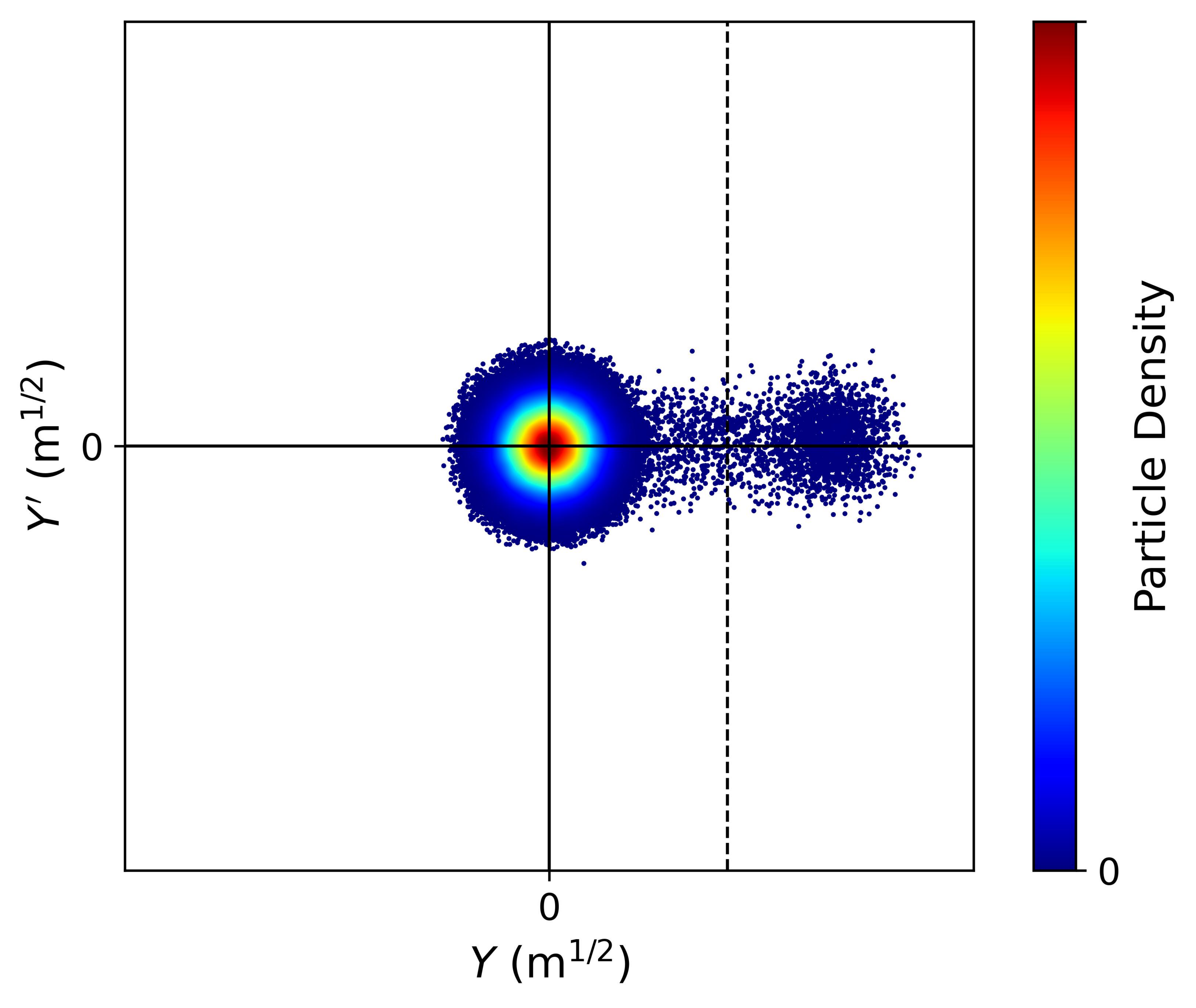}}
	\caption{Motion of particles extracted in flattop duration and falling edge in normalized phase space. (a) Initial particle distribution (befored kicked by the stripline kicker). (b) particles in the falttop duration and falling edge are kicked. (c) kicked particles move into the ESe (the vertical dashed line represents the $y$ coordinate of the ESe).}
	\label{fig:TOP_FALL}
\end{figure*}

The motion of particles extracted in the flattop duration and falling edge in the normalized phase space ($Y-Y^{\prime}$) is illustrated in Fig.~\ref{fig:TOP_FALL}.
The distribution of particles between the main bunch and the extracted bunch in the falttop duration in Fig.~\ref{fig:TOP_FALL_ca_ese_in_85} is form by the overlap of different kick amplitude in the falling edge.

\subsection{Feedback method for Controlling the Spot Dose}\label{Sec.2.B}
If the kicker pulse time is given, the extracted particle number can be calculated by Eqs.~\eqref{eq:Flattop_num} and \eqref{eq:N_fall_expression}.
On the contrary, for a given scanning spot requiring $N_{\mathrm{ext}}$ protons, the kicker pulse time can also be determined through these two equations:

First, choose a proper and fixed fall time $\tau_{\mathrm{fall}}$ of the kick pulse and obtain the discrete form of $\lambda\left(\tau\right)$ through BCM measurement.
Second, find $\tau_1 = n_1\cdot\Delta\tau$ using Eq.~\eqref{eq:Flattop_Begin}.
Then move $\tau_3$ and $\tau_2 = \tau_3 - \tau_{\mathrm{fall}}$ step by step and calculate the theoretically extracted number of particles for each step using Eqs.~\eqref{eq:Flattop_num} and \eqref{eq:N_fall_expression}.
The step length of the movement is decided by the resolution of $\lambda\left(\tau\right)$ measurement.
Finally, stop movement if the theoretically extracted particle number reach the desired $N_{\mathrm{ext}}$.
The detailed algorithm is described in Alg.~\ref{alg:Kicker_Pulse_Time_Determination_Process}.

\subsection{Beam Loss at Septum}\label{Sec.2.C}
In conventional fast extraction, no beam loss occurs at the septum because the kicker field rises and falls only in the gaps between bunches.
In the longitudinal localized kick driven fast extraction method, however, particles are present during the entire kicker pulse, including its rise/fall times.
Although the rising edge is aligned with a vacant longitudinal segment, the falling edge will inevitably overlap with the populated part of the bunch.
Particles experiencing this falling edge will receive an insufficient kick angle.
This leads to the distribution in the normalized phase space cut by the septum, as shown in Fig.~\ref{fig:TOP_FALL_ca_ese_in_85}.
\begin{figure}[!htb]
	\centering
	\includegraphics*[width=0.99\columnwidth]{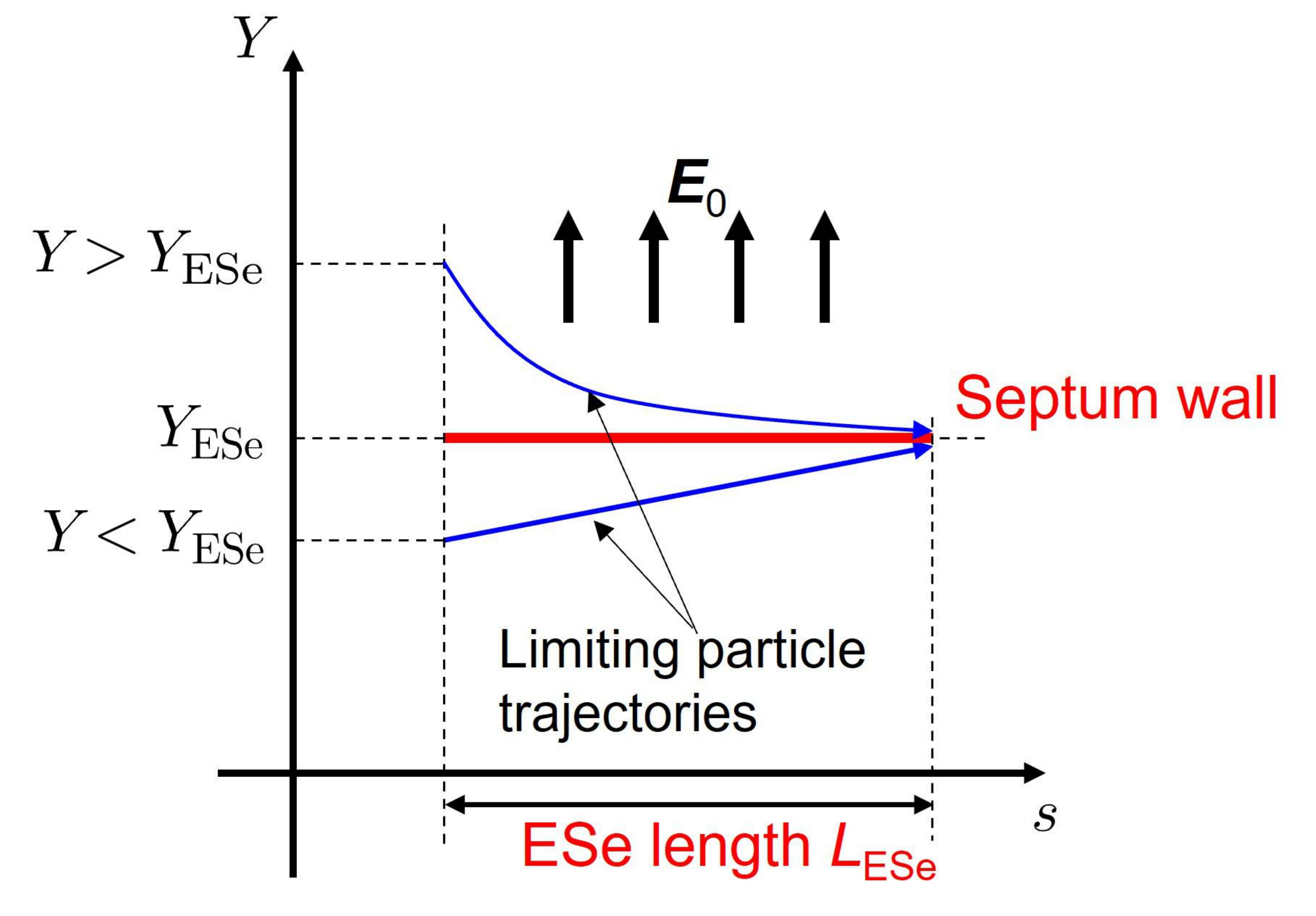}
	\caption{Particle losses on the ESe due to insufficient kick strength at the falling edge.}
	\label{fig:Schematic_of_Beam_Loss_y-s}
\end{figure}

The beam loss is origined from particle distribution cut by the septum and Fig.~\ref{fig:Schematic_of_Beam_Loss_y-s} exhibits the particles motion trajectory in the $Y-s$ plane\cite{Badano_385378}.

For particle with coordinate $\left(Y,Y^\prime\right)$ satisfying $Y<Y_{\mathrm{ESe}}$, it will collide with the septum wall (wire) if
\begin{equation}\label{Loss-CL}
	Y^{\prime} \ge Y^{\prime}_{\mathrm{CL}} = \left(\alpha - \frac{\beta}{L_{\mathrm{ESe}}}\right)\left(Y - Y_{\mathrm{ESe}}\right) + Y^{\prime}_{\mathrm{ESe}}
\end{equation}
where $\alpha$ and $\beta$ is the optical functions of the synchrotron ring at the entrance of the ESe, and $L_\mathrm{ESe}$ is the length of ESe.
$\left( Y_{\mathrm{ESe}}, Y^{\prime}_{\mathrm{ESe}} \right)$ is the coordinate of the anode of the ESe at the entrance of the ESe.
The lost particles are called “Lost in the circulating beam pipe ($\mathbf{Loss-CL}$)”.

Typically the thickness of the septum wall is $\dif y_{\mathrm{ESe}} = 0.1 \mathrm{mm}$ and particles hitting the \SI{0.1}{mm} thick septum wall directly at the entrance of the ESe is called “Lost at the septum wall ($\mathbf{Loss-SW}$)”.

For particle with coordinate $Y > Y_{\mathrm{SW}} = Y_{\mathrm{ESe}} + \dif Y_{\mathrm{ESe}}$ ($\dif Y_{\mathrm{ESe}} = \dif y_{\mathrm{ESe}} / \sqrt{\beta}$ in the normalized phase space), judging weather it will collide with the septum wall is a little bit more complex:

if $Y_{\mathrm{SW}} \le Y \le Y_{\mathrm{EX}} = Y_{\mathrm{ESe}} + \dif Y_{\mathrm{ESe}} + \Delta\theta L_{\mathrm{ESe}}/\left(2\sqrt{\beta}\right)$, the particles loss judgement is
\begin{equation}\label{Loss-EX-1}
	Y^{\prime} \le Y^{\prime}_{\mathrm{EX,1}} = \alpha\left(Y - Y_{\mathrm{ESe}}\right) + Y^{\prime}_{\mathrm{ESe}} - \sqrt{2\frac{\beta\Delta\theta}{L_{\mathrm{ESe}}} \sqrt{\beta} \left(Y - Y_{\mathrm{SW}}\right)}
\end{equation}
where $\Delta\theta$ is the deflecting angle of the ESe.

And for $Y \ge Y_{\mathrm{EX}}$, the particle loss judgement is then
\begin{widetext}
	\begin{equation}\label{Loss-EX-2}
		Y^{\prime} \le Y^{\prime}_{\mathrm{EX,2}} = \left(\alpha - \frac{\beta}{L_{\mathrm{ESe}}}\right)\left(Y - Y_{\mathrm{SW}}\right) + Y^{\prime}_{\mathrm{ESe}} + \alpha\dif Y_{\mathrm{ESe}}  - \frac{\sqrt{\beta}\Delta\theta}{2}
	\end{equation}
\end{widetext}
The particles satisfying $Y \ge Y_{\mathrm{SW}}$ has entered the ESe to be extracted, so the lost perticles are called  “Lost in the extraction channel ($\mathbf{Loss-EX}$)”.

Since the distribution cut by the ESe is overlaped by the particles kicked by different amplitude in the falling time, and particles with the same kick have an isotropic Gaussian distribution in the mormalized phase space, so the overlaped particle number distribution $\rho$ can be written as
\begin{equation}\label{eq:fall_dist}
	\rho\left(Y,Y^{\prime}\right) = \frac{\lambda_{Y}}{\sqrt{2\pi}\sigma}\mathrm{exp}\left(-\frac{Y'^2}{2\sigma^2}\right)
\end{equation}
indicating that the distribution in the $Y$ direction is uniform with line density of $\lambda_{Y}$ and the distribution in the $Y^{\prime}$ direction is Gaussian: $N\left(0,\sigma^2\right)$.

\begin{figure}[!htb]
	\centering
	\subfloat[\label{fig:Lose_Region_ALL}]
	{\includegraphics*[width=0.45\linewidth]{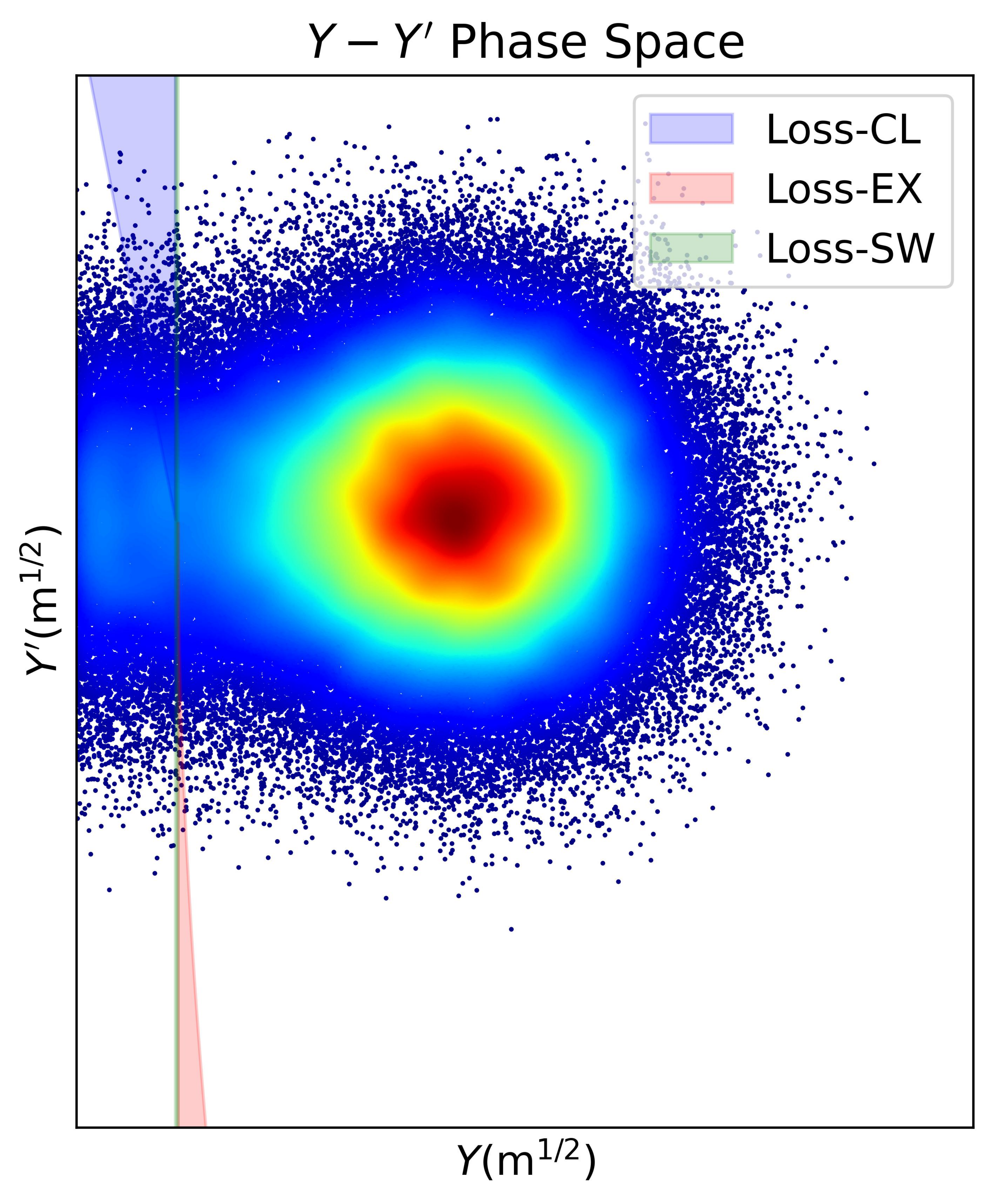}}
	\subfloat[\label{fig:Lose_Region_ZOOM}]
	{\includegraphics*[width=0.55\linewidth]{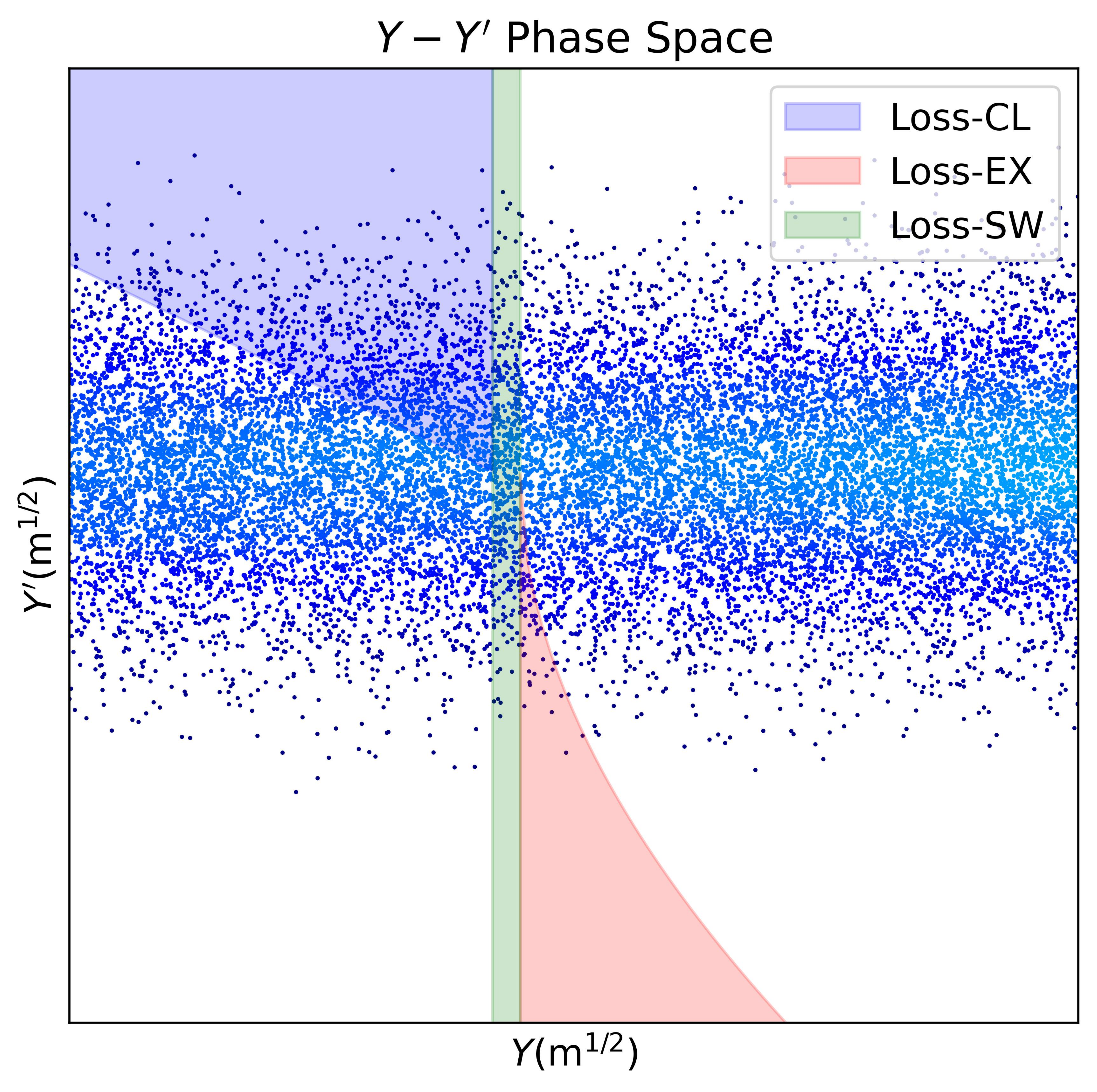}}
	\caption{The distribution of lost particles in normalized phase space. (a) Lost region compared to all of the kicked particles in the entrance of ESe. (b) zoom of (a).}
	\label{fig:3_Loss}
\end{figure}

Therefore, the lost number of particles can be theoretically calculated:
\begin{equation}\label{loss_num_CL}
	\begin{aligned}
		N_{\mathrm{Loss-CL}} &= \int_{-\infty}^{Y_{\mathrm{ESe}}}\int_{Y^{\prime}_{\mathrm{CL}}}^{+\infty} \rho\left(Y,Y^{\prime}\right) \dif Y^{\prime} \dif Y \\
		&= \frac{\lambda_{Y}\sigma}{\left(\frac{\beta}{L_{\mathrm{ESe}}}-\alpha\right)\sqrt{2\pi}}
	\end{aligned}
\end{equation}
\begin{equation}\label{loss_num_SW}
	N_{\mathrm{Loss-SW}} = \int_{Y_{\mathrm{ESe}}}^{Y_{\mathrm{SW}}} \rho\left(Y,Y^{\prime}\right) \dif Y = \lambda_{Y}\dif Y_{\mathrm{ESe}}
\end{equation}
\begin{equation}\label{loss_num_EX}
	\begin{aligned}
		N_{\mathrm{Loss-EX}} =& \int_{Y_{\mathrm{SW}}}^{Y_{\mathrm{EX}}}\int_{-\infty}^{Y^{\prime}_{\mathrm{EX,1}}} \rho\left(Y,Y^{\prime}\right) \dif Y^{\prime} \dif Y \\
		&+ \int_{Y_{\mathrm{EX}}}^{+\infty}\int_{-\infty}^{Y^{\prime}_{\mathrm{EX,2}}} \rho\left(Y,Y^{\prime}\right) \dif Y^{\prime} \dif Y
	\end{aligned}
\end{equation}
The integral of $N_{\mathrm{Loss-EX}}$ has no elementary expression and can only be solved by numerical integration.
Fig.~\ref{fig:3_Loss} shows these 3 parts of beam loss in the normalized phase space.

For each scanning spot, the extracted particles $N_{\mathrm{ext}}$ consists of two parts: during flattop $N_{\mathrm{flattop}}$ and falling time $N_{\mathrm{fall}}$.
In the assumption of uniform falling edge line density $\lambda_{\mathrm{fall}}$ and fixed falling edge width $\tau_{\mathrm{fall}}$, $N_{\mathrm{ext}}$ can be expressed as
\begin{equation}\label{N_ext}
	N_{\mathrm{ext}} = N_{\mathrm{flattop}} + k_{1}\cdot\lambda_{\mathrm{fall}}
\end{equation}
where $k_{1}$ is a constant, indicating the directly proportional relationship between $N_{\mathrm{fall}}$ and $\lambda_{\mathrm{fall}}$ according to Eq.~\eqref{eq:N_fall_expression}.

As mentioned before, $\lambda_{Y}$ is produced by the overlap of a series of Gaussian distribution, so under these assumption of uniform $\lambda_{\mathrm{fall}}$ and fixed $\tau_{\mathrm{fall}}$, it can be derived that $\lambda_{Y}$ is directly proportional to $\lambda_{\mathrm{fall}}$: $\lambda_{Y} \propto \lambda_{\mathrm{fall}}$.

Then according to Eqs.~\eqref{loss_num_CL}, \eqref{loss_num_SW} and \eqref{loss_num_EX}, we have directly proportional relationship
\begin{equation}\label{N_loss}
	N_{\mathrm{Loss}} \propto \lambda_{Y}, \lambda_{Y} \propto \lambda_{\mathrm{fall}} \Rightarrow N_{\mathrm{Loss}} = k_{2}\cdot\lambda_{\mathrm{fall}}
\end{equation}
where $k_{2}$ is also a constant.

So the particle losing rate can be written as
\begin{equation}\label{eq:lose_rate}
	\frac{N_{\mathrm{Loss}}}{N_{\mathrm{ext}}} = \frac{k_{2}}{k_{1}+\frac{N_{\mathrm{flattop}}}{\lambda_{\mathrm{fall}}}}
\end{equation}
indicating that increased $\lambda_{\mathrm{fall}}$ will result in increased particle losing rate.
This conclusion is qualitative since the condition of uniform $\lambda_{\mathrm{fall}}$ is generally not satisfied.
But the basic dependency relationship between losing rate and line density $\lambda_{\mathrm{fall}}$ is confident and can be validated by the simulation results presented in Sec.~\ref{Sec.4.B}.

\section{RCS Lattice Design}\label{Sec.3}
\subsection{General Design Parameters}\label{Sec.3.A}
The magnetic field of the bending magnets in the RCS is described by
\begin{equation}\label{eq:BM_curve}
	B\left( t \right) = B_{0} - B_{1}\cos\left( 2\pi ft \right)
\end{equation}
where $B_{0} = \SI{0.5528}{T}$, $B_{1} = \SI{0.4024}{T}$ and $f = \SI{25}{Hz}$, yielding a maximum field $B_\mathrm{max} = \SI{0.9552}{T}$.
The maximum bending field in the RCS is generally lower than that of typical medical slow-cycling synchrotrons (approximately $\SI{1.5}{T}$).
This difference is primarily due to eddy current effects, which can cause temperature rise and power loss.
By optimizing the bending magnet design, including the pole-end shape, core-end slots, coil material, and lamination thickness, a field following Eq.~\eqref{eq:BM_curve} can be achieved.
For instance, the bending field can be ramped from $\SI{0.16}{T}$ to $\SI{0.98}{T}$ within $\SI{20}{ms}$ in CSNS RCS\cite{6068232}.

Other design parameters of the RCS are listed in Table~\ref{tab:RCS_design_parameters}:
\begin{table}[!htb]
	\caption{\label{tab:RCS_design_parameters}
		Design parameters of the RCS
	}
	\begin{ruledtabular}
		\begin{tabular}{lll}
			parameter & value & unit\\
			\colrule
			particle type                & proton            & -          \\
			inject energy                & 7                 & MeV/u      \\
			inject magnetic rigidity     & 0.38              & T$\cdot$ m \\
			extract energy               & 70-250            & MeV/u      \\
			extract magnetic rigidity    & 1.23-2.43         & T$\cdot$ m \\
			stored particles per cycle   & $2\times10^{11}$  & -          \\
			maximum bending field        & $\sim1.0$         & T          \\
		\end{tabular}
	\end{ruledtabular}
\end{table}
Based on the maximum extraction magnetic rigidity and the maximum bending field, the total length of the bending magnets is calculated as \SI{16.0}{m}.
Therefore, 8 bending magnets (BMs) are adopted, each with a length of \SI{2}{m}, corresponding to a bending radius of \SI{2.55}{m}.

\subsection{Fast Extraction System Requirements}\label{Sec.3.B}
\begin{figure*}[!htb]
	\centering
	\includegraphics*[width=1.80\columnwidth]{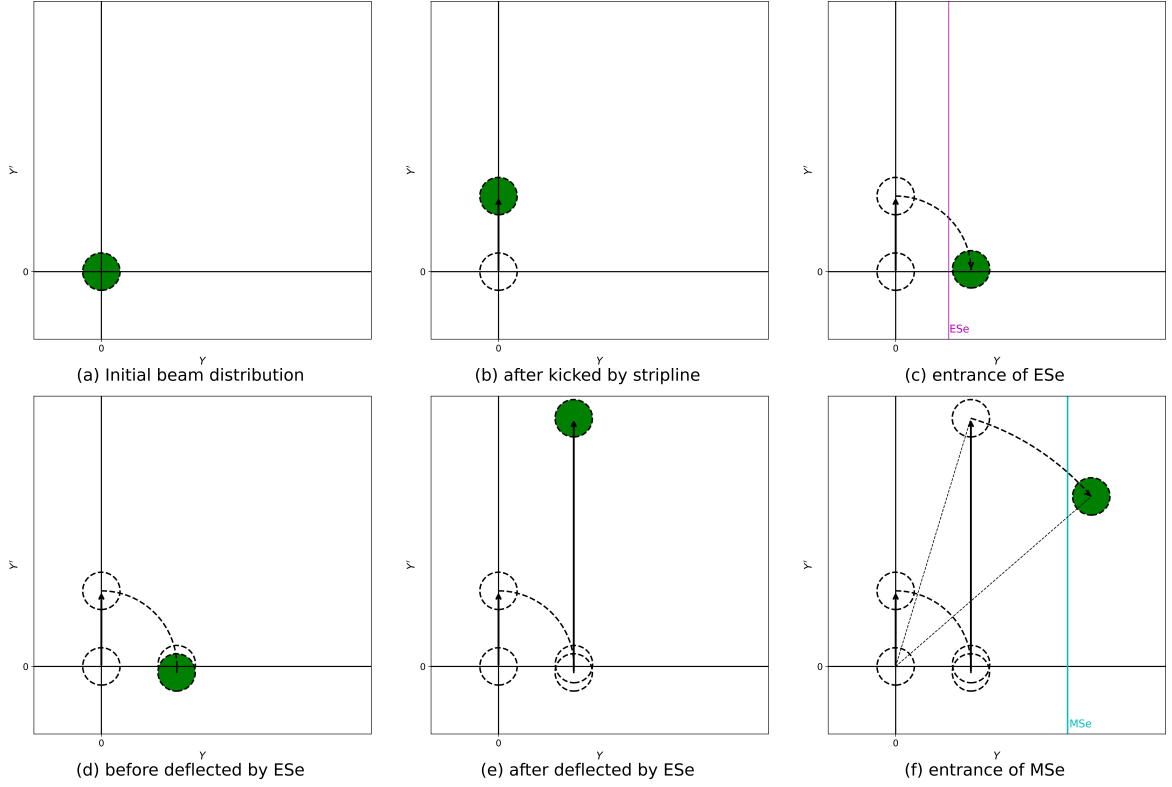}
	\caption{Extracted proton bunch motion in normalized phase space. The circle with dashed edge represents the bunch. Th egreen solid circle is the bunch's curent position and the hollow circle is the past position to show the dynamic motion of the bunch. Different figures mean different position in the synchrotron ring. (a) Initial position before kicked by the stripline kicker; (b) after kicked by the stripline kicker; (c) at the entrance of the ESe; (d) at the center of the ESe and befored deflected bu it; (e) after deflected by the ESe; (f) at the entrance of the MSe.}
	\label{fig:Kicker_ESe_MSe}
\end{figure*}

The fast extraction system consisting of a stripline kicker, an ESe, and an MSe has been exhibited in Fig.~\ref{fig:Fast_Extraction_Layout}.
The motion of the extracted bunch in normalized phase space is depicted in Fig.~\ref{fig:Kicker_ESe_MSe}.
For a deflecting element of length $L$ with a uniform field along its length, the element can be treated as a zero-length element located at its center.
This equivalence has been applied in the deflecting element of stripline kicker and ESe in Fig.~\ref{fig:Kicker_ESe_MSe}.

If the beam receives a kick at the center of the stripline kicker and propagates to the ESe, the resulting separation at the entrance of the ESe, $\Delta y_{\mathrm{ESe,in}}$, can be expressed as
\begin{equation}\label{eq:Sep_ES_in}
	\Delta y_{\mathrm{ESe,in}} = \theta_{\mathrm{kicker}}\cdot\sqrt{\beta_{\mathrm{kicker}}\beta_{\mathrm{ESe,in}}}\sin{\mu_{1}}
\end{equation}
where $\theta_{\mathrm{kicker}}$ is the kick angle imparted by the stripline kicker, $\beta_{\mathrm{kicker}}$ and $\beta_{\mathrm{ESe,in}}$ are the $\beta$ functions at the center of the kicker and at the entrance of the ESe, and $\mu_{1}$ is the phase advance between these two locations.

Similarly, after further deflection by the ESe, the separation at the entrance of the MSe, $\Delta y_{\mathrm{MSe,in}}$, is given by
\begin{equation}\label{eq:Sep_MS_in}
	\begin{aligned}
		\Delta y_{\mathrm{MSe,in}}=&\ \theta_{\mathrm{kicker}}\cdot\sqrt{\beta_{\mathrm{kicker}}\beta_{\mathrm{MSe,in}}}\sin{\left(\mu_{1}+\dif\mu+\mu_{2}\right)} \\
		&+\theta_{\mathrm{ESe}}\cdot\sqrt{\beta_{\mathrm{ESe}}\beta_{\mathrm{MSe,in}}}\sin{\mu_{2}}
	\end{aligned}
\end{equation}
where $\theta_{\mathrm{ESe}}$ is the kick angle of the ESe, $\beta_{\mathrm{ESe}}$ and $\beta_{\mathrm{MSe,in}}$ are the $\beta$ functions at the center of the ESe and at the entrance of MSe, respectively, $\mu_{2}$ is the phase advance between the center of the ESe and the entrance of the MSe, and $\dif\mu$ is the phase advance from the entrance to the center of the ESe.

The basic principle of the longitudinal localized kick driven fast extraction system is to achieve the desired transverse beam separation distance with a kick amplitude that is technically feasible with current technology.
This means the $Y$ coordinate of the bunch must be greater than that of ESe in Fig.~\ref{fig:Kicker_ESe_MSe}(c) and MSe Fig.~\ref{fig:Kicker_ESe_MSe}(f).

\begin{figure*}[!htb]
	\centering
	\subfloat[\label{fig:TOP_FALL_noREVERSEKICK_ca_kicker_1_in_85}]
	{\includegraphics*[width=0.30\linewidth]{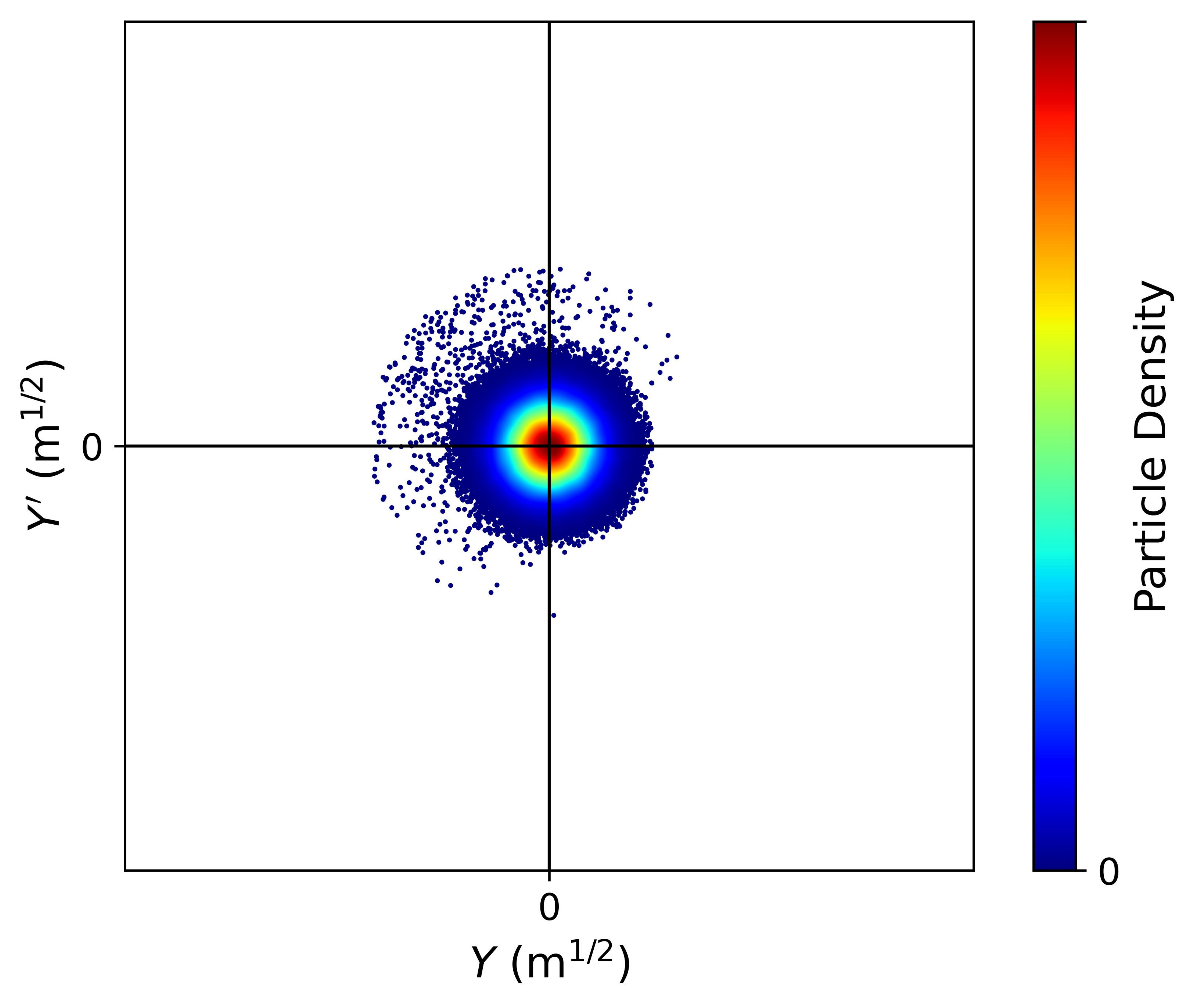}}
	\subfloat[\label{fig:TOP_FALL_noREVERSEKICK_ca_kicker_1_out_85}]
	{\includegraphics*[width=0.30\linewidth]{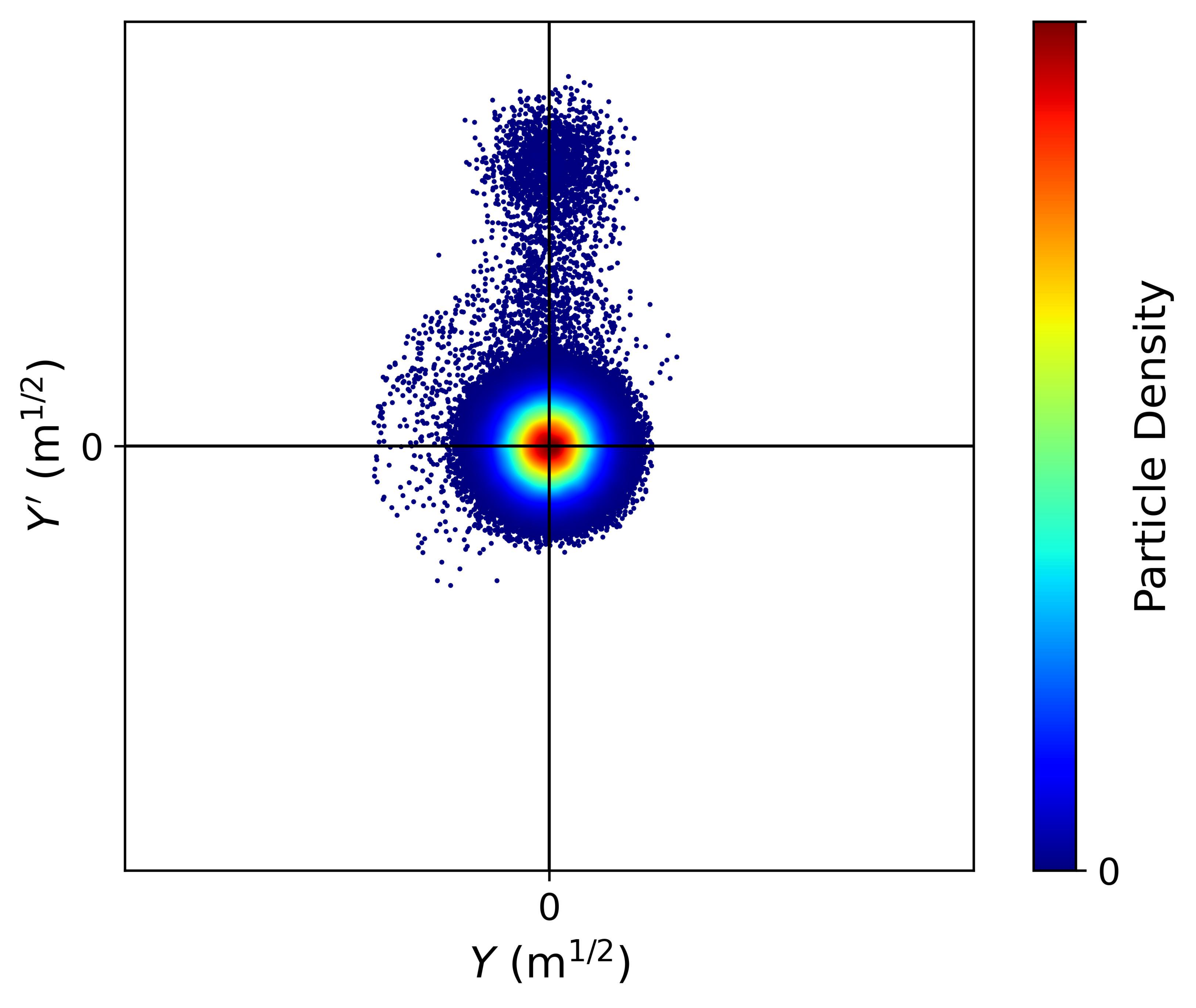}}
	\subfloat[\label{fig:TOP_FALL_noREVERSEKICK_ca_ese_in_85}]
	{\includegraphics*[width=0.30\linewidth]{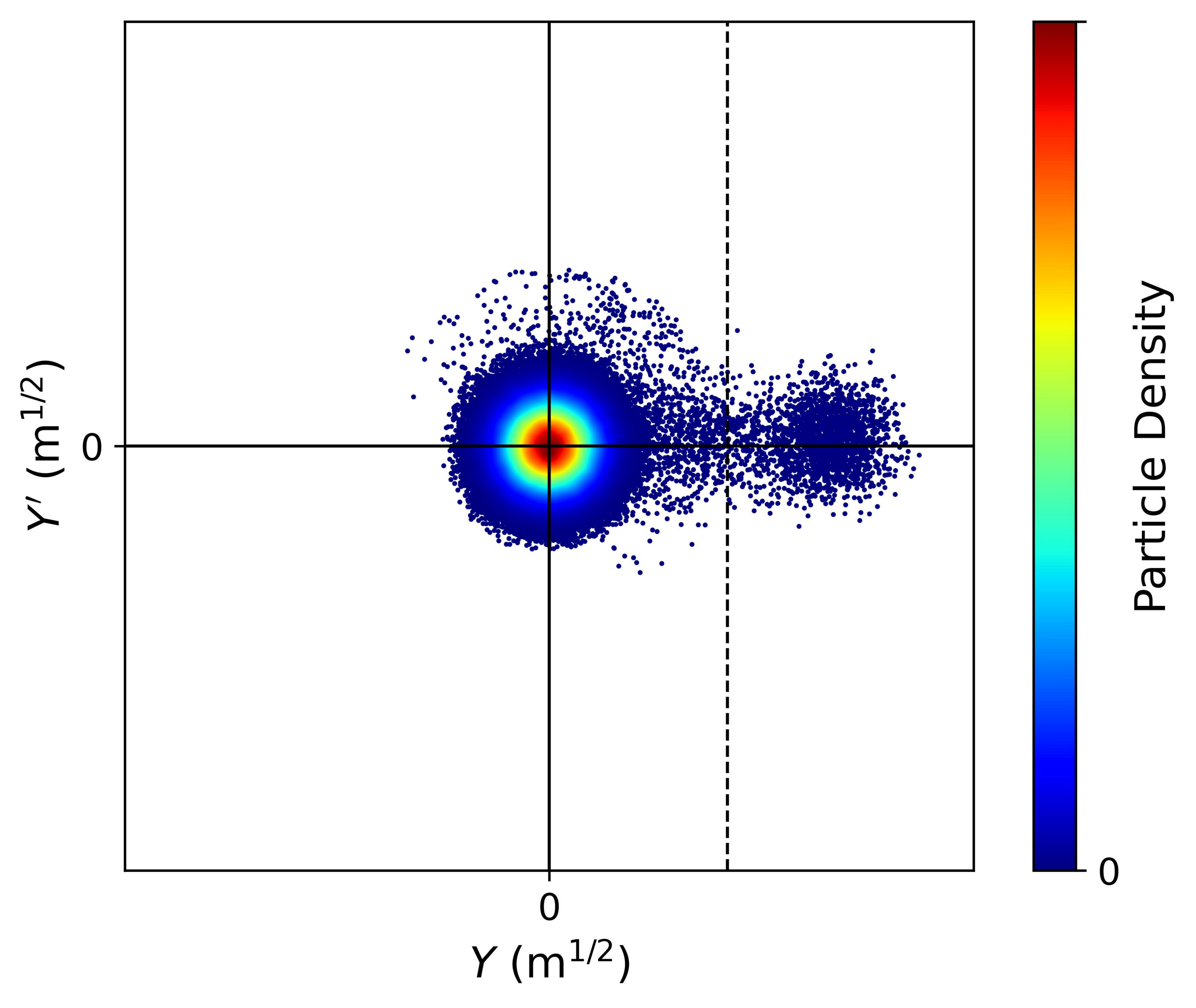}}
	\caption{Motion of particles with non-Gaussian distribution in normalized phase space. (a) befored kicked by the stripline kicker. (b)after kicked by the stripline kicker. (c) kicked particles move into the ESe (the vertical dashed line represents the $y$ coordinate of the ESe).}
	\label{fig:TOP_FALL_noREVERSEKICK}
\end{figure*}

Another consideration arises from the falling edge of the stripline kicker pulse.
Fig.~\ref{fig:Kicker_ESe_MSe} displays the motion of proton bunch only in the flattop duration.
The motion of particles in the falling edge is shown in Fig.~\ref{fig:TOP_FALL}.
In Fig.~\ref{fig:TOP_FALL_ca_ese_in_85}, particles in the left of the dashed line will not be extracted and continue to move inside the ring.
This movement will make the particle distribution non‑Gaussian, as illustrated in Fig.~\ref{fig:TOP_FALL_noREVERSEKICK}.
The non-Gaussian distribution makes Eq.~\eqref{eq:N_fall_expression} calculating number of particles extracted in the fall time no longer accurate, thus affecting the spot dose calculation.

\begin{figure*}[!htb]
	\centering
	\subfloat[\label{fig:TOP_FALL_REVERSEKICK_ca_ese_in_85}]
	{\includegraphics*[width=0.30\linewidth]{Figures/TOP_FALL_ca_ese_in_85}}
	\subfloat[\label{fig:TOP_FALL_REVERSEKICK_ca_kicker_2_in_85}]
	{\includegraphics*[width=0.30\linewidth]{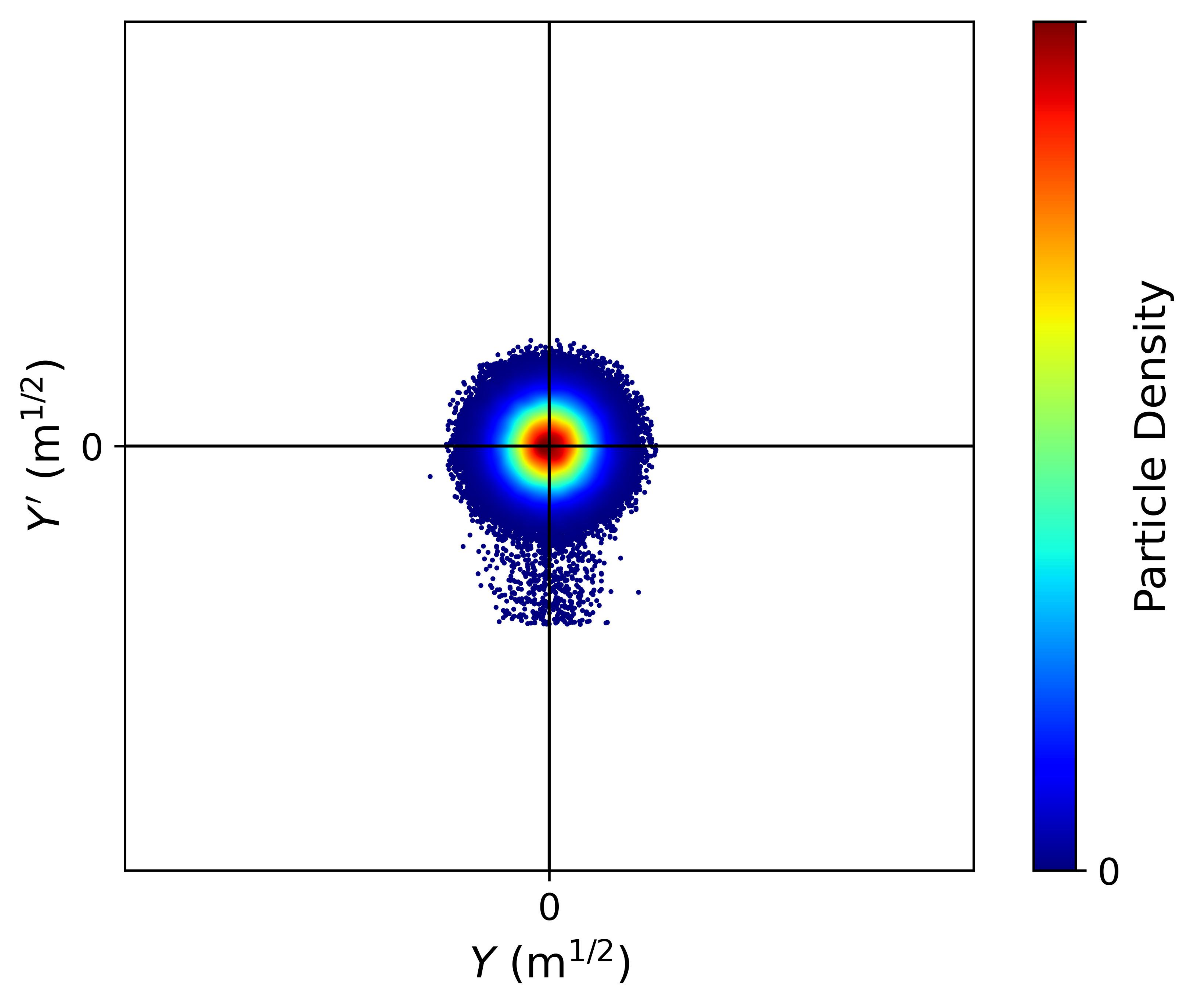}}
	\subfloat[\label{fig:TOP_FALL_REVERSEKICK_ca_kicker_2_out_85}]
	{\includegraphics*[width=0.30\linewidth]{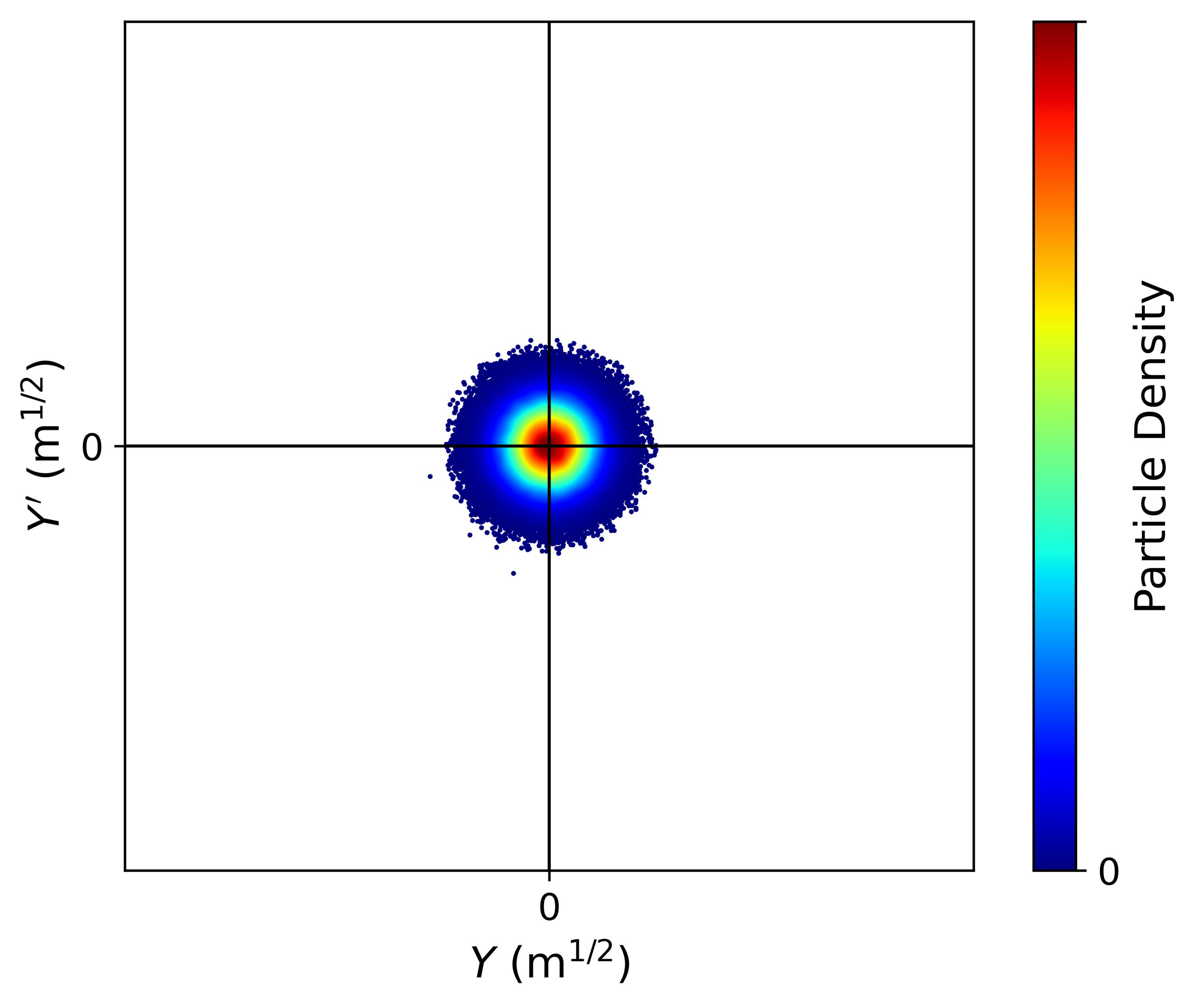}}
	\caption{Schematic of reverse kick to recover Gaussian distribution in the normalized phase space. (a) at the entrance of ESe (Fig.~\ref{fig:TOP_FALL_ca_ese_in_85}). (b) $180\deg$ phase advance in relative to the stripline kicker in Fig.~\ref{fig:TOP_FALL_ca_kicker_1_out_85}. (c) particle distribution recoverd to Gaussian after the reverse kick.}
	\label{fig:Reverse_Kick}
\end{figure*}

To mitigate this effect, the unextracted particles during the falling edge must return to the main bunch in the normalized phase space.
The solution is to apply a reverse kick with the same magnitude but opposite direction, as depicted in Fig.~\ref{fig:Reverse_Kick}.
The reverse kick is also realized by a stripline kicker.
Comparing Fig.~\ref{fig:TOP_FALL_ca_kicker_1_out_85} and Fig.~\ref{fig:TOP_FALL_REVERSEKICK_ca_kicker_2_in_85}, it is clear that these two stripline kickers shoule be designed to have the same pulse time and voltage, but with a phase advance of $180\deg$.


\subsection{RCS Lattice Design Results}\label{Sec.3.C}
\begin{figure}[!htb]
	\includegraphics*[width=0.99\columnwidth]{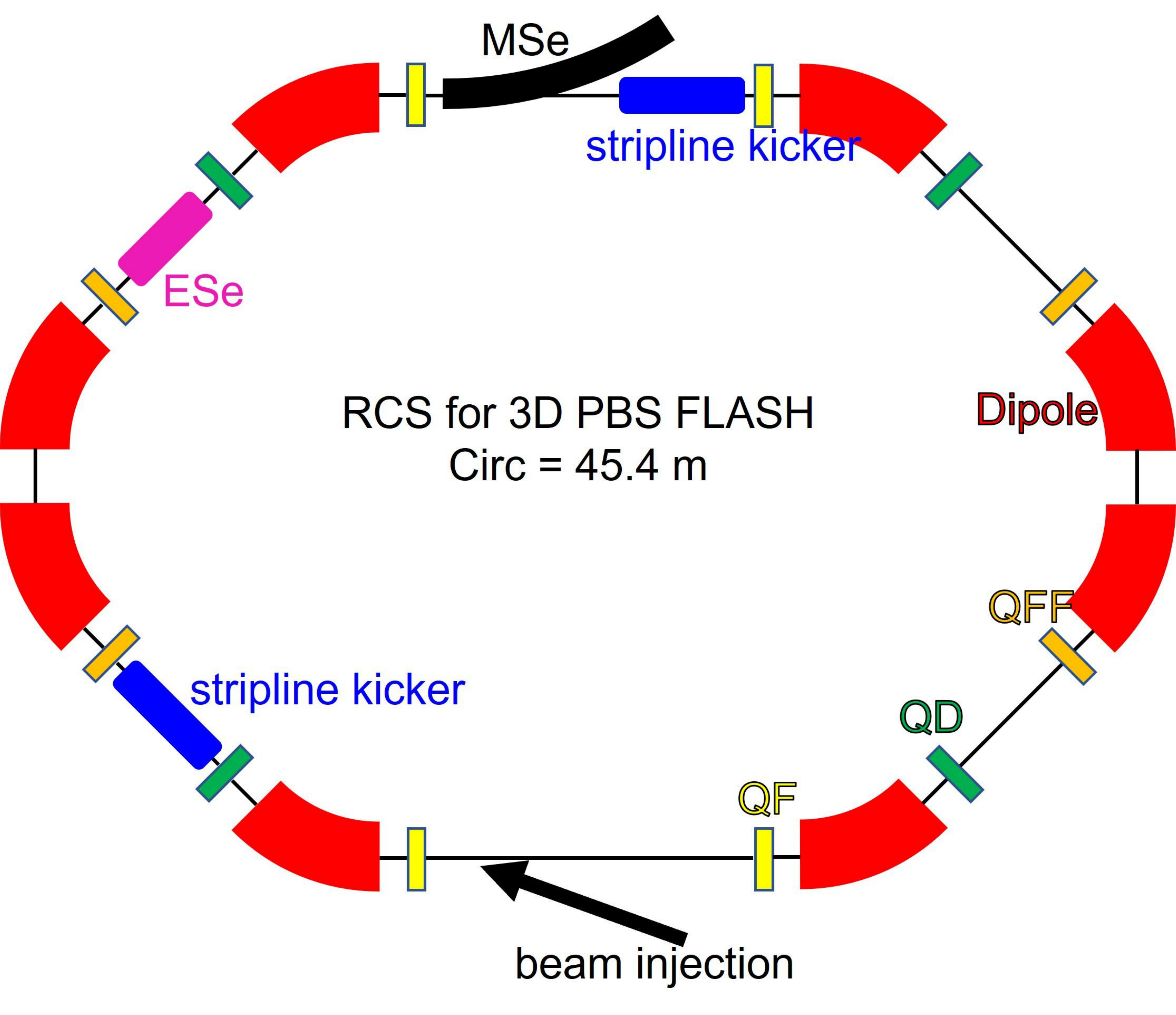}
	\caption{\label{fig:RCS_Lattice}Layout of the RCS.}
\end{figure}

\begin{figure}[!htb]
	\includegraphics*[width=0.99\columnwidth]{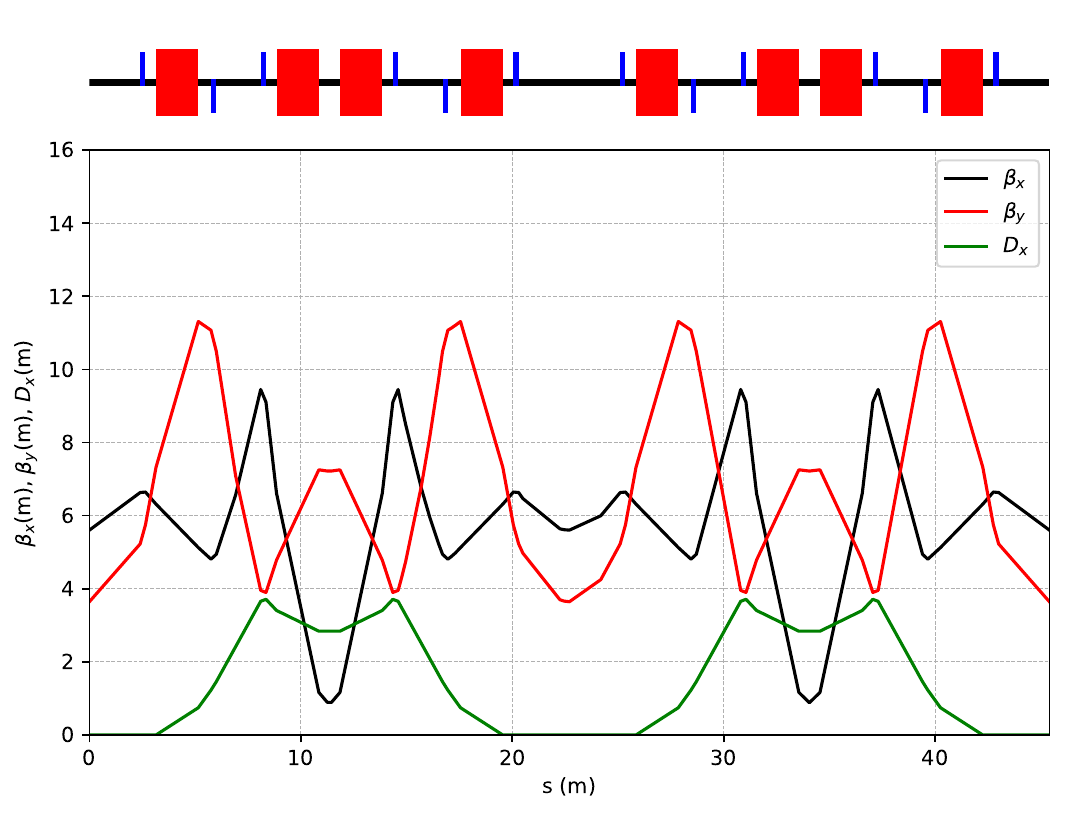}
	\caption{\label{fig:RCS_twiss}Twiss parameters of the RCS.}
\end{figure}

Following the design specifications outlined in Sec.~\ref{Sec.3.A} and Sec.~\ref{Sec.3.B}, the RCS lattice was designed with the layout shown in Fig.~\ref{fig:RCS_Lattice} and the Twiss parameters presented in Fig.~\ref{fig:RCS_twiss}.
Key parameters of the synchrotron are summarized in Table~\ref{tab:RCS_key_parameters}.
\begin{table}[!htb]
	\caption{\label{tab:RCS_key_parameters}
		Key parameters of the RCS
	}
	\begin{ruledtabular}
		\begin{tabular}{lll}
			parameter & value & unit\\
			\colrule
			Circumference                & 45.4          & m          \\
			Transition Energy            & 2.059         & -          \\
			Maximum $\beta_{x,y}$        & 9.45/11.3     & m          \\
			Maximum desperation $D_{x}$  & 3.711         & m          \\
			Tune($x/y$)                  & 1.680/1.230   & -          \\
			Natural chromaticity($x/y$)  & -0.491/-2.545 & -          \\
			Edge angle of BM             & 18            & deg        \\
		\end{tabular}
	\end{ruledtabular}
\end{table}

Compared with existing medical RCS designs (Table~\ref{tab:RCS_design}), the circumference in our design is larger.
This results primarily from limiting the maximum bending magnet field to below \SI{1.0}{T}, which necessitates longer total bending magnet length.
Furthermore, the present design must satisfy specific phase advance requirements for the longitudinal localized kick driven fast extraction system.
A longer circumference provides greater flexibility in arranging the necessary elements.
The phase advance between the center of the stripline kicker and the entrance of the ESe is $\mu_{1} = 88.23\deg$, the phase advance between the entrance and the center of the ESe is $\dif\mu = 6.31\deg$, and the phase advance between the center of the ESe and the entrance of the MSe is $\mu_{2} = 32.16\deg$.
The particle motion illustrated in Fig.~\ref{fig:Kicker_ESe_MSe} of Sec.~\ref{Sec.3.B} corresponds precisely to the extracted bunch in normalized phase space at an energy of \SI{70}{MeV}.

Next, we will calculate the parameters of the two primary deflecting elements: the stripline kicker and the ESe.

\begin{figure}[!htb]
	\centering
	\subfloat[\label{fig:DeflectionAngle_kicker_vs_Ek}]
	{\includegraphics*[width=0.49\linewidth]{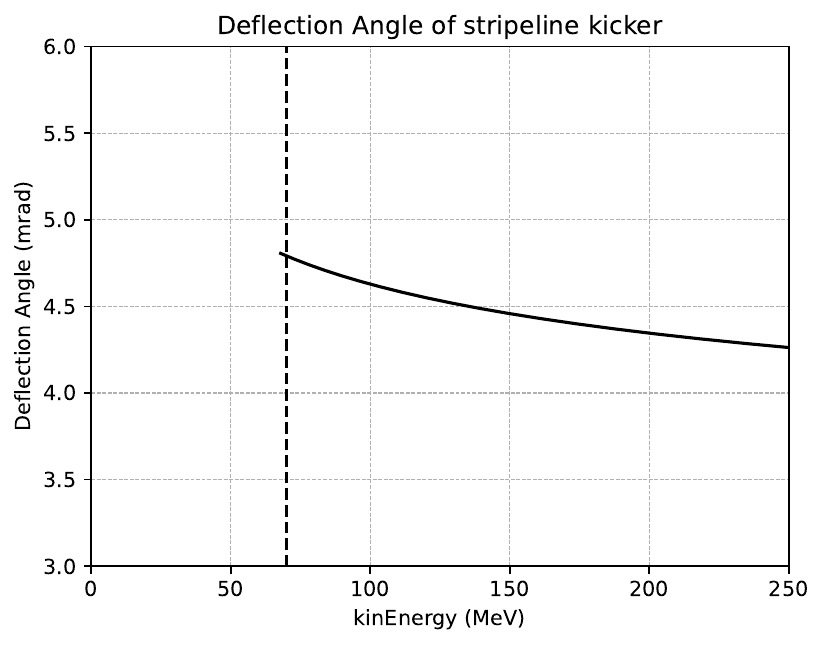}}
	\subfloat[\label{fig:kickerVoltage_vs_Ek}]
	{\includegraphics*[width=0.49\linewidth]{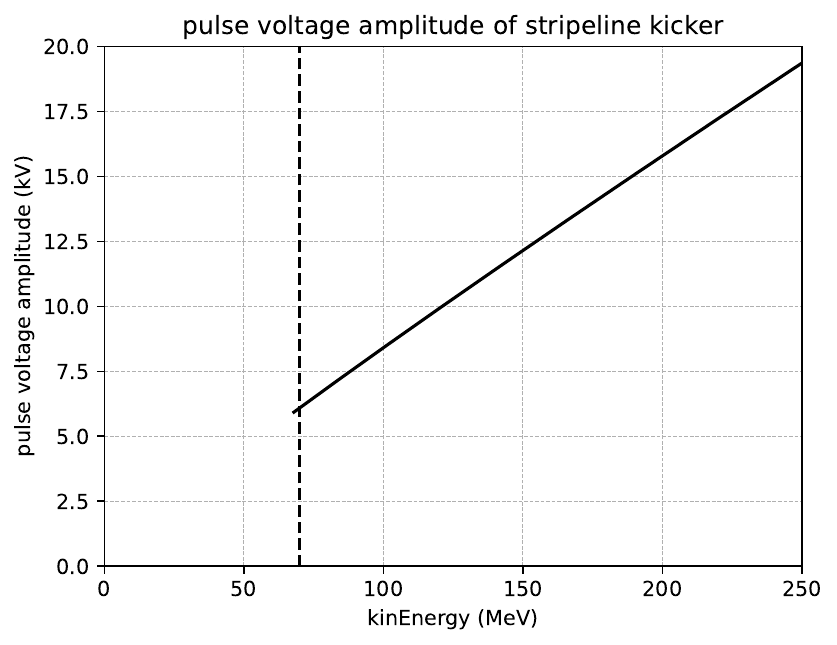}}
	\caption{Deflection angle (a) and pulse voltage amplitude (b) of the stripline kicker with respect to the beam energy.}
	\label{fig:kicker-angle-voltage}
\end{figure}

Considering an existing \SI{15}{mm} extraction orbit bump at the entrance of ESe generated by 3 bump magnets, the resulting deflection angle as a function of beam energy is shown in Fig.~\ref{fig:DeflectionAngle_kicker_vs_Ek}. 
To avoid beam loss from particles hitting the electrodes, the electrode gap is chosen as $d=\SI{60}{mm}$.
The stripline length and width are set to $L=\SI{1.8}{m}$ and $w=\SI{60}{mm}$, respectively.
The corresponding required pulse voltage amplitude versus beam energy is presented in Fig.~\ref{fig:kickerVoltage_vs_Ek}.
The maximum voltage reaches \SI{19.36}{kV}, which is technically feasible with existing technology.

\begin{figure}[!htb]
	\centering
	\subfloat[\label{fig:DeflectionAngle_ESe_vs_Ek}]
	{\includegraphics*[width=0.49\linewidth]{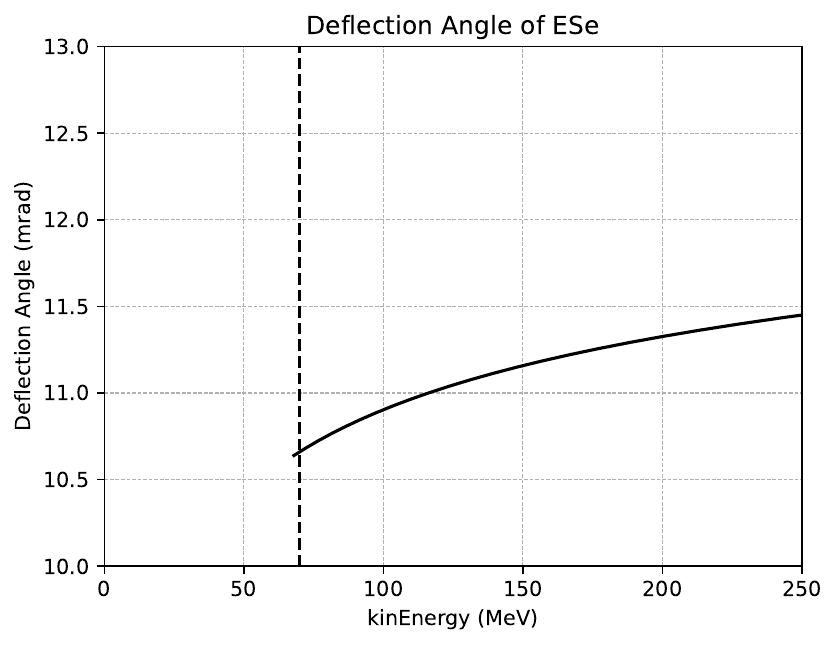}}
	\subfloat[\label{fig:ESeVoltage_vs_Ek}]
	{\includegraphics*[width=0.49\linewidth]{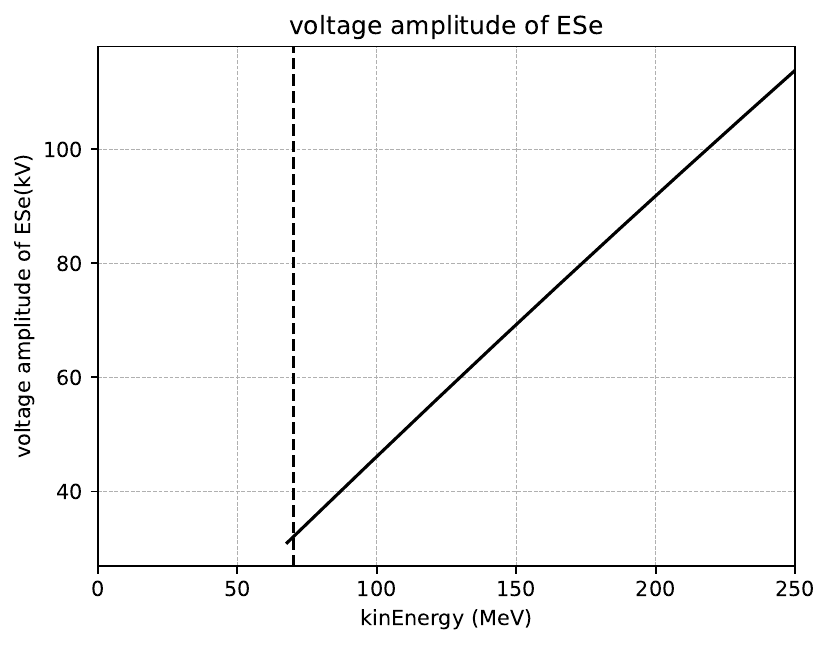}}
	\caption{Deflection angle (a) and pulse voltage amplitude (b) of the ESe with respect to the beam energy.}
	\label{fig:ESe-angle-voltage}
\end{figure}

For the ESe, the required separation at the entrance of the MSe is \SI{56.8}{mm}.
Given an ESe length of \SI{1.35}{m} and an electrode gap of \SI{30}{mm}, its deflection angle and pulse voltage amplitude as functions of beam energy are shown in Fig.~\ref{fig:ESe-angle-voltage}.
The maximum electrode voltage required is \SI{113.84}{kV}, which still remains within achievable limits.


\section{Simulation Results}\label{Sec.4}
To validate the feasibility of the longitudinal localized kick driven fast extraction method, we performed simulations using “SynTrack”—a high-speed, parallel particle tracking code.
Derived from Li-Track\cite{Yao:IPAC2019-WEPTS033}, SynTrack is a high-performance program designed for synchrotron beam dynamics simulations.
Based on C++, it employs a symplectic integration algorithm to ensure physical accuracy in long-term tracking.
A feedback module based on Alg.~\ref{alg:Kicker_Pulse_Time_Determination_Process} has been developed and implemented.
A module with the function of applying a reverse kick with the same magnitude but opposite direction is also added to the code to finish the simulation.

In the simulation, the time resolution of the BCM was set to \SI{0.1}{ns}.
For the falling edge, a fixed width of achievable \SI{3}{ns} was adopted\cite{LIU2020163670}, corresponding to the fastest achievable falling speed of the stripline kicker.
A simple plan is set in which the spots are delivered with a uniform \SI{5}{mm} spot spacing and uniform monitor units (MU) per spot.
The plan is designed to deliver dose from \SI{4}{cm} depth (proton energy \SI{70}{MeV}) and $2\times10^{8}$ particles is requirement to enter ESe in each fast extraction process.
In this paper, the number of particles entering the ESe is regared as extracted number of particles.

\subsection{Results of Spot Dose Error}
For convenience, we use spot dose error to describe the accuracy here (higher spot dose error is equivalent to lower accuracy):
\begin{equation}\label{eq:error_def}
	\mathrm{spot\ dose\ error} = \frac{N_{\mathrm{ext,simulation}} - N_{\mathrm{ext,design}}}{N_{\mathrm{ext,design}}} \times 100\%
\end{equation}

\begin{figure}[!htb]
	\centering
	\includegraphics*[width=0.90\columnwidth]{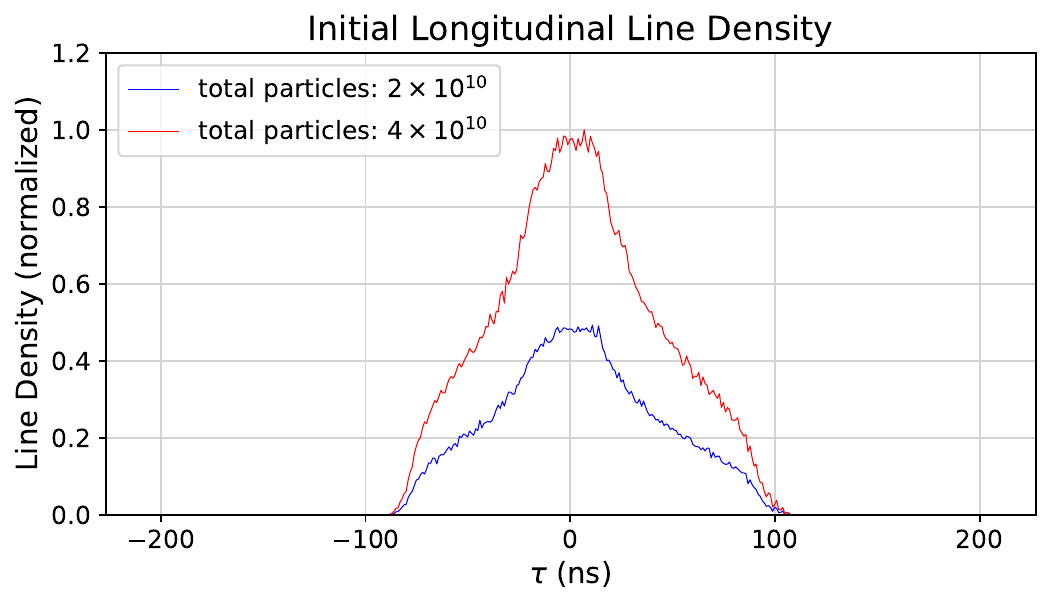}
	\caption{Comparison of the longitudinal line density distributions for the two simulated cases.}
	\label{fig:Longitudinal_Line_Density}
\end{figure}

Simulations were conducted under two conditions: one with a total of $2\times10^{10}$ particles and the other with $4\times10^{10}$ particles sotred in the ring before extraction, which can provide around 100 and 200 scanning spots, respectively.
The longitudinal distribution range was identical in both settings, implying a two-fold difference in line density $\lambda\left(\tau\right)$, as shown in Fig.~\ref{fig:Longitudinal_Line_Density}.

\begin{figure}[!htb]
	\centering
	\includegraphics*[width=0.90\columnwidth]{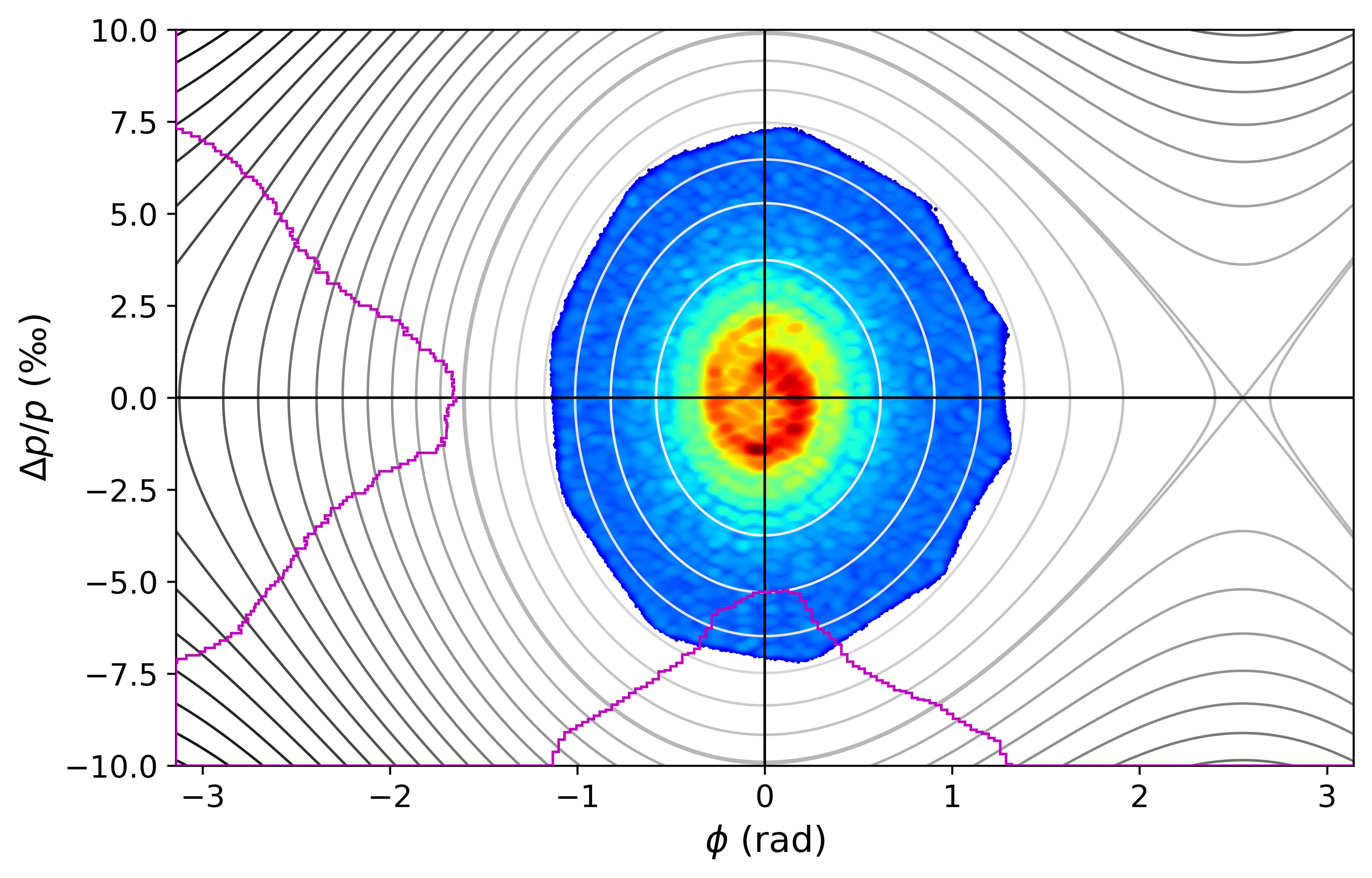}
	\caption{Longitudinal phase space at a certain turn in the extraction process.}
	\label{fig:Longitudinal_Phase_Space}
\end{figure}

The longitudinal phase space at a certain turn in the extraction process is shown in Fig.~\ref{fig:Longitudinal_Phase_Space}.
The phase space illustrates that the process of longitudinal localized fast extraction is like gradually ‘peeling off’ particles from the outside to the inside.

\begin{figure*}[!htb]
	\centering
	\subfloat[$2\times10^{10}$ total particles with \SI{0}{ns} pulse fall time\label{fig:Spill_Structure_TOP_20e+04}]
	{\includegraphics*[width=0.33\linewidth]{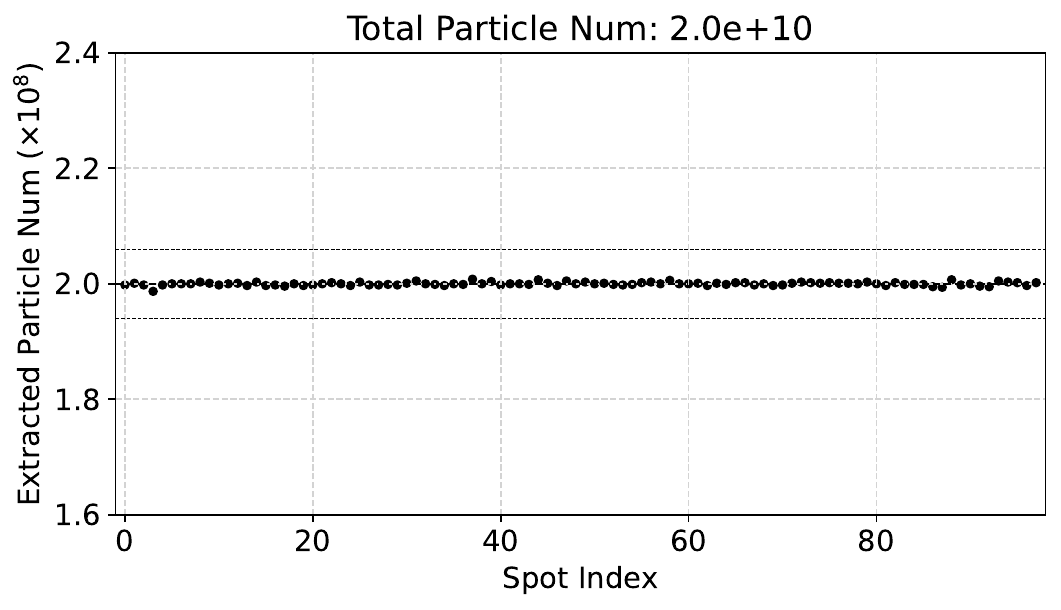}}
	\subfloat[$2\times10^{10}$ total particles with \SI{3}{ns} pulse fall time\label{fig:Spill_Structure_TOP_FALL_20e+04}]
	{\includegraphics*[width=0.33\linewidth]{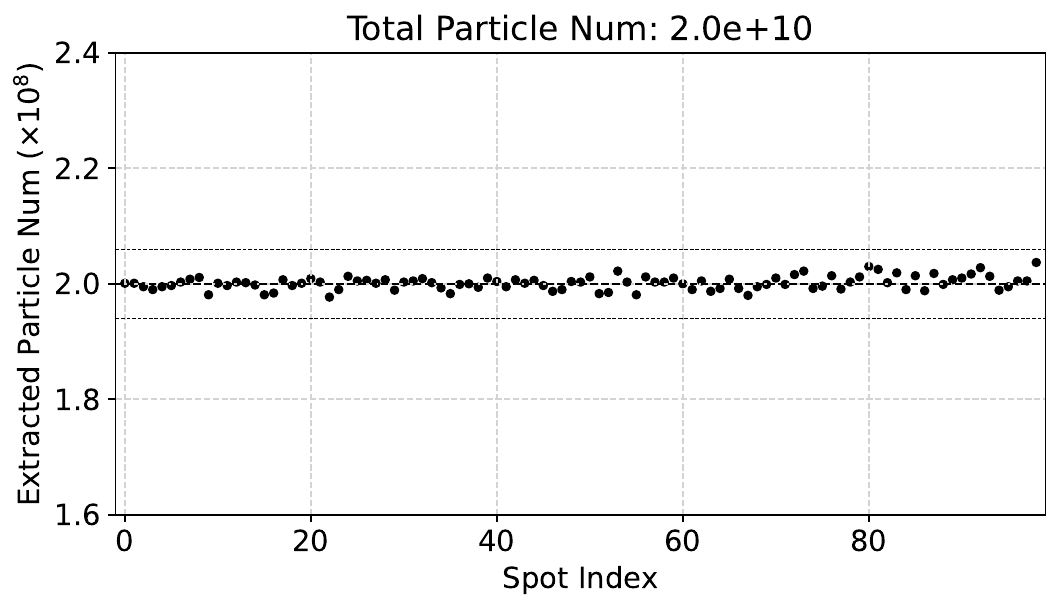}}
	\subfloat[$4\times10^{10}$ total particles with \SI{3}{ns} pulse fall time\label{fig:Spill_Structure_TOP_FALL_40e+04}]
	{\includegraphics*[width=0.33\linewidth]{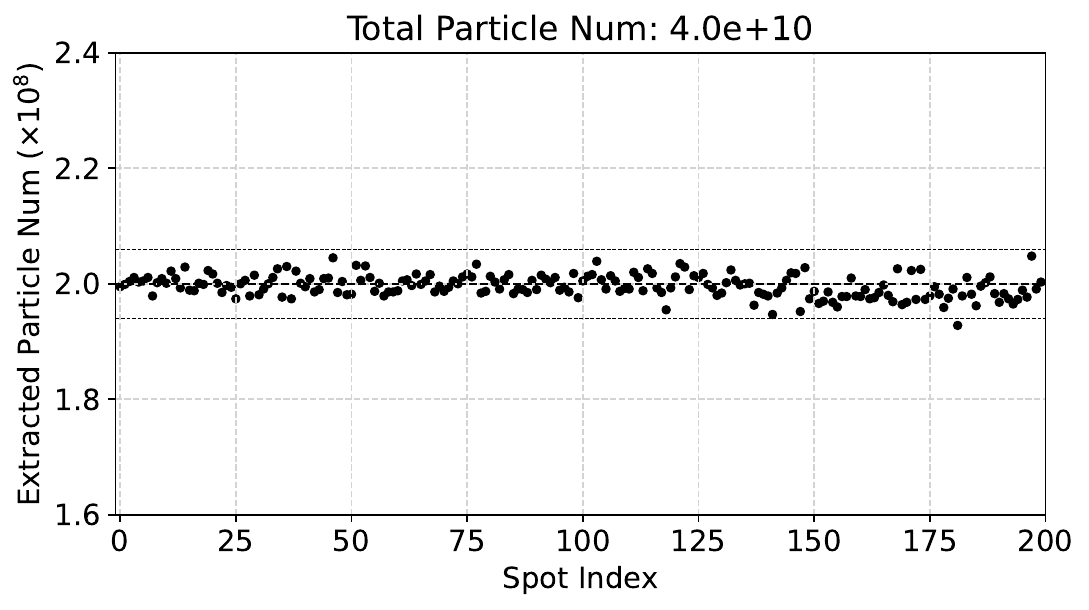}}
	\caption{Extracted number of particles at each scanning spot with a total of $2\times10^{10}$ and $4\times10^{10}$ particles stored and different fall time. The three horizontal dashed lines are the designed number of particles to be extracted ($2\times10^{8}$), and the clinical acceptable error of $\pm3\%$ for medical use. The \SI{0}{ns} pulse fall time in (a) is an ideal situation just for comparison with (b).}
	\label{fig:Spill_Structure}
\end{figure*}

The number of particles extracted for each scanning spot is presented in Fig.~\ref{fig:Spill_Structure}.

The design criterion is to make sure more than 95\% of the scanning spots have dose error less than 3\%\cite{https://doi.org/10.1002/mp.13622}.
It is obvious that the 3 cases in Fig.~\ref{fig:Spill_Structure} all satisfy this criterion.
The comparison between Fig.~\ref{fig:Spill_Structure_TOP_20e+04} and Fig.~\ref{fig:Spill_Structure_TOP_FALL_20e+04} shows that the error is mainly resulted from the existance of falling edge.

\begin{figure}[!htb]
	\centering
	\includegraphics*[width=0.90\columnwidth]{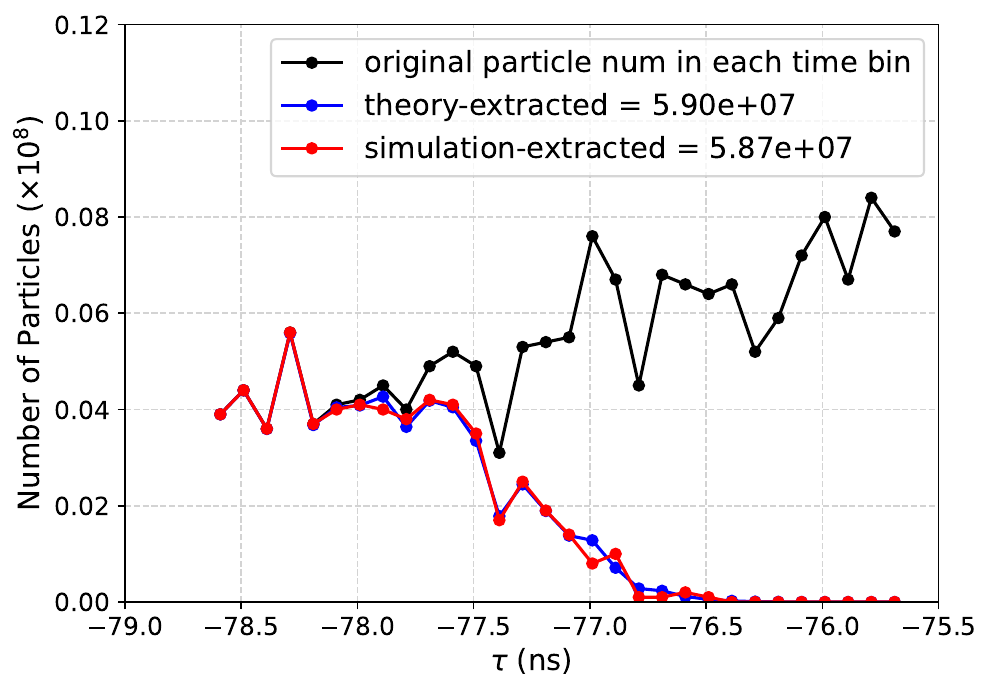}
	\caption{A comparison of theoretical and simulated values for particle extraction per time bin during the falling edge. The black line indicates the initial particle population in each bin. The blue line corresponds to the theoretically targeted number of particles for extraction, while the red line shows the number of particles actually extracted in the simulation.}
	\label{fig:FALL_CAL_vs_SIM}
\end{figure}

The reduction of spot dose accuracy can be attributed to 2 factors: one is the deviations of the real beam distribution from an ideal transverse Gaussian profile, which is an inherent systematic errors in the extraction process.
The other is the theorrtically designed $N_{\mathrm{fall,design}}$ in the simulation is calculated discretely in the code.
That is, the sum of 30 time bins (\SI{3}{ns} fall time with \SI{0.1}{ns} resolution) are sumed to calculate the continuous integration of Eq.~\eqref{eq:N_fall_expression}.
Fig.~\ref{fig:FALL_CAL_vs_SIM} compares the calculated and simulated numbers of particles extracted in each bin during the pulse fall time.

\begin{figure}[!htb]
	\centering
	\includegraphics*[width=0.90\columnwidth]{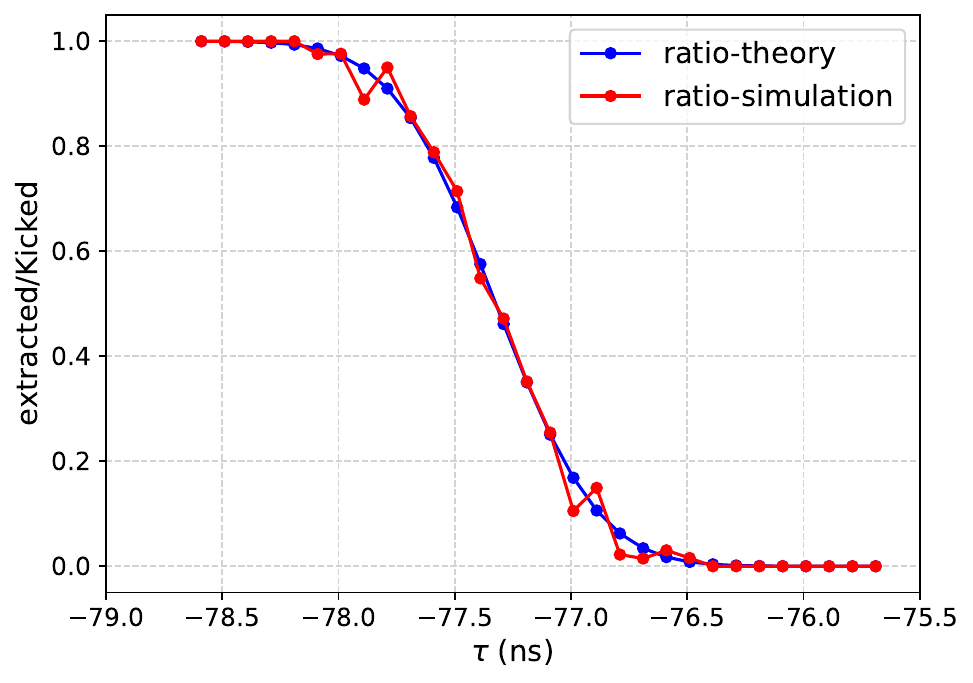}
	\caption{The ratio of particles actually and theoretically extracted.}
	\label{fig:RATIO_CAL_vs_SIM}
\end{figure}

The ratio of extracted particles in simulation and in theory calculation is shown in Fig.~\ref{fig:RATIO_CAL_vs_SIM} (red line).
The blue line in Fig.~\ref{fig:RATIO_CAL_vs_SIM} corresponds to the Complementary Error Function part
\begin{equation}\label{eq:erfc}
	\frac{1}{2} \mathrm{erfc}\left( \frac{Y_{\mathrm{ESe}}-\frac{\tau_3-\tau}{\tau_3-\tau_2}Y_{c,\mathrm{fall}}}{\sqrt{2\varepsilon_{\mathrm{rms}}}} \right)
\end{equation}
in Eq.~\eqref{eq:N_fall_expression}.
Fig.~\ref{fig:FALL_CAL_vs_SIM} and Fig.~\ref{fig:RATIO_CAL_vs_SIM} validate the effectiveness of the Eq.~\eqref{eq:N_fall_expression}.

The comparison between Fig.~\ref{fig:Spill_Structure_TOP_FALL_20e+04} and Fig.~\ref{fig:Spill_Structure_TOP_FALL_40e+04} shows that increased total particle number will result in increased spot dose error.
As analized before, the error mainly arises from the pulse fall time, but the discrepancy between theory and simulation is unpredictable because both the particle number and the transverse particle distribution vary at each time bin for every scaning spot.
Therefore, we conduct a qualitative analysis here:
Considering that the error in the falttop duration is negligible, the definition of spot dose error in Eq.~\eqref{eq:error_def} can be rewritten as
\begin{equation}\label{eq:error_def_fall}
	\begin{aligned}
		\mathrm{spot\ dose\ error} &= \frac{N_{\mathrm{fall,simulation}} - N_{\mathrm{fall,design}}}{N_{\mathrm{ext,design}}} \times 100\% \\
		&= \frac{\Delta N_{\mathrm{fall}}}{N_{\mathrm{ext,design}}} \times 100\%
	\end{aligned}
\end{equation}
The two simulation settings have the same longitudinal distribution range, but different total particles means different line density in the falling edge ($\lambda_{\mathrm{fall}}$).
In the condition of fixed fall time of \SI{3}{ns}, simulation setting with larger line density have more particles extracted during the pulse fall time ($N_{\mathrm{fall}}$).
And the deviation of extracted in the falling edge $\Delta N_{\mathrm{fall}}$ will also increase, causing the spot does error to be larger.

\subsection{Error Analysis}
In addition to the inherent error introduced by the Gaussian assumption and discrete integration in the pulse fall time, errors from the hardware and measurements of the fast extraction system should also be carefully considered.
Sec.~\ref{Sec.3.A} has presented the theoretical calculation formular of extracted particles, and the error type can be determined from the formulas:
\begin{itemize}
	\item The error of line density ($\lambda\left(\tau\right)$) measured by BCM, Gaussian with $\sigma = 0.5\%$.
	\item The error of closed orbit (affecting $Y_{c,\mathrm{fall}}$), fixed value of $\pm 0.05\ \mathrm{mm}$.
	\item The error of the rms transverse emittance ($\varepsilon_{\mathrm{rms}}$), Gaussian with $\sigma = 5.0\%$.
	\item The error of the stripline kicker pulse voltage, Gaussian with $\sigma = 0.5\%$.
	\item The error of the stripline kicker pulse time, Gaussian with $\sigma = 0.06\ \mathrm{ns}$.
\end{itemize}
The choice to treat the closed orbit error as a fixed value is motivated by the need to optimize computational resource usage during simulation.

The analysis of error from fast extraction system is also conducted in the two settings of $2\times10^{10}$ and $4\times10^{10}$ total particles.
The simulation parameters are kept unchanged from those used in the error-free case and the results of error analysis are listed in Table.~\ref{tab:error_analysis}.
\begin{table*}[!htb]
	\caption{\label{tab:error_analysis}Simulation results of error analysis}
	\begin{ruledtabular}
		\begin{tabular}{llrr}
			\multirow{2}{*}{total particles} & \multirow{2}{*}{error value} & \multicolumn{2}{c}{95\% max error of spot dose} \\
			& & $2.0\times10^{10}$ & $4.0\times10^{10}$ \\
			\hline
			Error-free & & 1.15\% & 1.90\% \\
			\hline
			Error of line density ($\lambda\left(\tau\right)$)                   & $\sigma = 0.50\%$   & 1.15\% & 1.95\% \\
			\multirow{2}{*}{error of closed orbit (CO)} & CO error = $+0.05$ mm & 1.20\% & 1.85\% \\
			                                            & CO error = $-0.05$ mm & 1.10\% & 2.10\% \\
			Error of the rms transverse emittance ($\varepsilon_{\mathrm{rms}}$) & $\sigma = 5.00\%$     & 1.20\% & 1.85\% \\
			Error of the stripline kicker pulse voltage                          & $\sigma = 0.50\%$   & 1.15\% & 1.85\% \\
			Error of the stripline kicker pulse time                             & $\sigma = 0.06$ ns & 3.25\% & 5.25\% \\
			\hline
			\multirow{2}{*}{All error} & CO error = $+0.05$ mm & 2.95\% & 6.00\% \\
			                           & CO error = $-0.05$ mm & 2.85\% & 5.90\% \\
		\end{tabular}
	\end{ruledtabular}
\end{table*}
The error value is chosen according to current techniques can reach\cite{Sterbini:IPAC2014-THPME173,Yao:NAPAC2016-WEPOB24,PASJ2016MOP110,PhysRevSTAB.14.051002}.
It is shown that most of the error types will not remarkably affect the spot dose accuracy, except for the error of  the stripline kicker pulse time.
This result suggests that the kicker time must be carefully controlled to reduce its affect.

When it comes to all types of error involved, simulation results indicate that the case of $4\times10^{10}$ total particles will no longer satisfy the criterion of 95\% scanning spots have dose error less than 3\%.
However, $2\times10^{10}$ total particles can still meet this criterion.

\subsection{Results of Beam Loss at Septum}\label{Sec.4.B}
\begin{figure}[!htb]
	\includegraphics*[width=0.99\columnwidth]{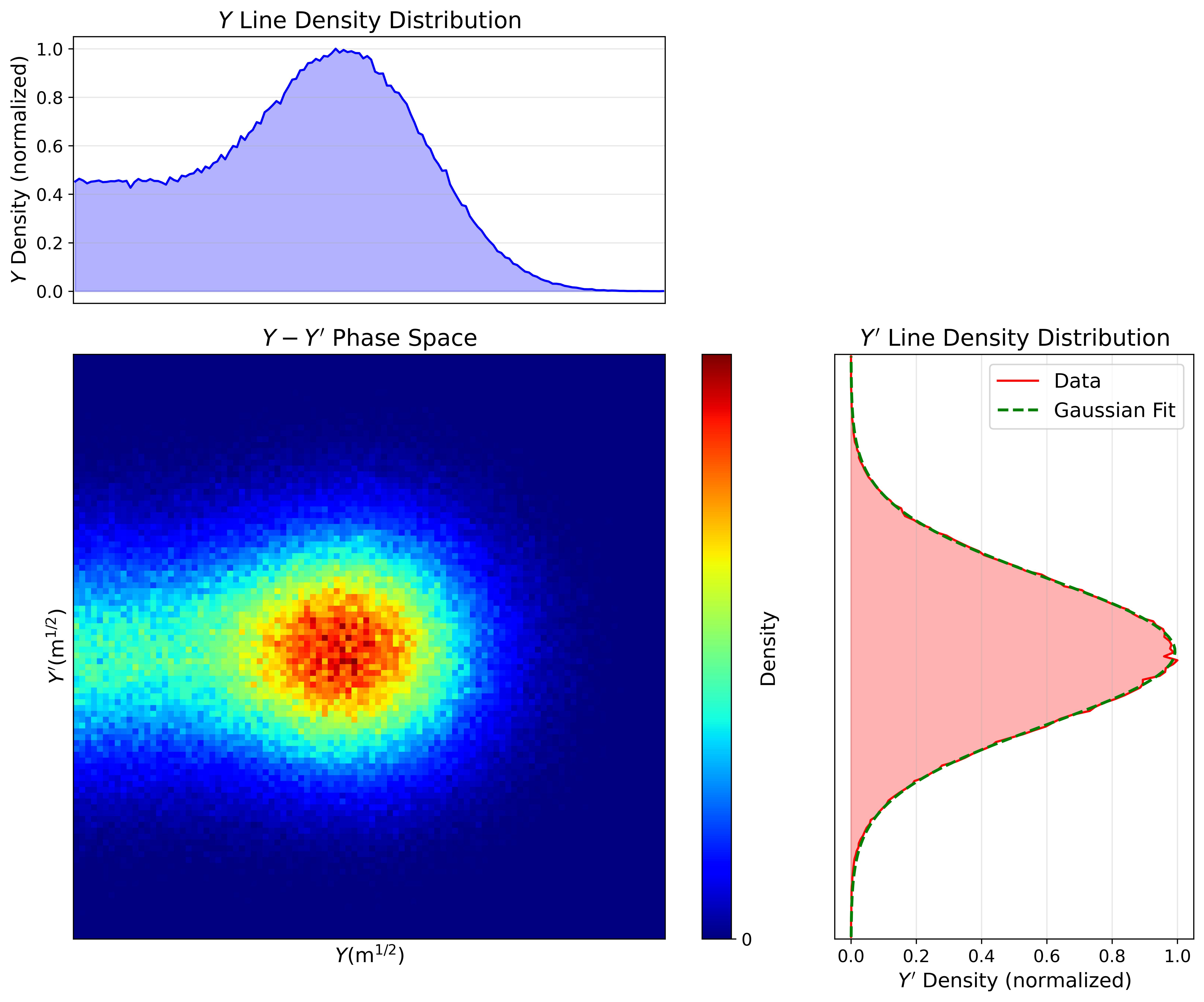}
	\caption{\label{fig:Lose_Region_ALL_FIT} Distribution of extracted particles in the $Y-Y^{\prime}$ phase space. The distribution is the overlap of all spots.}
\end{figure}

In sec.~\ref{Sec.2.C}, the beam loss introduced by the septum has been analysed and calculated theroetically.
Fig.~\ref{fig:Lose_Region_ALL_FIT} shows the simulation results of extracted particle distribution.
The line density of $Y$ and $Y^{\prime}$ coordinates varify that the assumption of uniform distribution in $Y$ direction and Gaussian distribution in $Y^{\prime}$ direction in Eq.~\eqref{eq:fall_dist} is reasonable.

The beam loss attributable to the fall time is summarized in Table~\ref{tab:septum_beam_loss} (the definition of “\textbf{Loss-EX}”, “\textbf{Loss-SW}” and “\textbf{Loss-CL}” is in Sec.~\ref{Sec.2.B}).
\begin{table*}
	\caption{\label{tab:septum_beam_loss}Beam Loss at the Septum}
	\begin{ruledtabular}
		\begin{tabular}{lllrrrrr}
			total particles                   & $\lambda_{Y}$ (fit value)  & $\sigma$ (fit value)     &        & \textbf{Loss-CL} & \textbf{Loss-SW} & \textbf{Loss-EX} & total Loss \\
			\hline
			\multirow{3}{*}{$2\times10^{10}$} & \multirow{3}{*}{$9.49\times10^{11}$} & \multirow{3}{*}{$1.58\times10^{-3}$} & theory & $1.15\times10^{8}$  & $4.15\times10^{7}$ & $6.05\times10^{6}$ & \\
			                                  & & & simulation         & $1.15\times10^{8}$ & $4.12\times10^{7}$ & $6.60\times10^{6}$ & 0.840\%\\
			                                  & & & deviation & 0.00\%             & -0.72\%            & 9.09\%             & \\
			\hline
			\multirow{3}{*}{$4\times10^{10}$} & \multirow{3}{*}{$3.20\times10^{12}$} & \multirow{3}{*}{$1.52\times10^{-3}$} & theory & $3.75\times10^{8}$  & $1.40\times10^{8}$ & $1.90\times10^{7}$ & \\
			                                  & & & simulation         & $3.77\times10^{8}$ & $1.41\times10^{8}$ & $2.07\times10^{7}$ & 1.377\%\\
			                                  & & & deviation & 0.53\%             & 0.71\%             & 8.95\%            & \\
		\end{tabular}
	\end{ruledtabular}
\end{table*}
The results show the theoretical and simulation beam loss match well, and the particles are mainly lost in the circulating beam pipe.
The comparison of total loss (last column in Table~\ref{tab:septum_beam_loss}) indicates that increased line density in the falling edge $\lambda_{\mathrm{fall}}$ will indeed cause the beam loss to be larger, which has been derived in Eq.~\eqref{eq:lose_rate}.

The beam loss level of around $1\%$ is an acceptable value and supports the applicability of the longitudinal localized kick driven fast extraction method to realize PBS FLASH delivery.

\section{Discussion}\label{Sec.5}
The longitudinal localized kick driven fast extraction method proposed in this paper is to fit the scanning method in which the beam is aligned parallel to the layer.
In this scanning pattern, the effective irradiation time only depends on the number of layers covered by the transverse beam size and the time for scanning each layer, but unrelated with the total layer number.
So ideally, the tumour volume target is unlimited.

However, the size of each scanning layer is limited by the number of particles stored in the RCS ring.
On the one hand, the space charge effect restricts the stored particles; on the other hand, the analysis and simulation results presented in this paper illustrate that to control the spot dose error, particle number can not be too much.
For a given plan with required dose and layering thickness, the size of the scanning layer is limited.
For instance, if \SI{8}{Gy} is needed to deliver to a layer with \SI{5}{mm} thickness, $2\times10^{10}$ total particles which can meet the spot dose accuracy criterion is only sufficient for around \SI{50}{cm^2} layer size.
If the particles need to supply 2 layers, only \SI{25}{cm^2} layer size is applicable.
This rough estimate does not take into account the loss during the beam transport process.
If this loss is considered, the layer size will be even smaller.

In cases involving tumors adjacent to organs at risk (OARs), a larger layer size may be necessary.
To enable this, the line density in the falling edge ($\lambda_{\mathrm{fall}}$) must be diluted before extraction.
This can be achieved by methods such as flattening the longitudinal phase space with multi-harmonic waves\cite{WOS:000813996500001} or blowing up the longitudinal emittance by adding phase noise to the RF wave\cite{BAUDRENGHIEN2013181}.
Additionally, since spatially fractionated radiation therapy (SFRT) can optimize dose distribution\cite{Prezado_2024,LI2024737}, combining the layer-parallel-to-beam scanning pattern with SFRT presents a potential solution for irradiating targets with large layer sizes.

\section{Conclusion}\label{Sec.6}
This paper presents and validates a longitudinal localized kick driven fast extraction method, supported by a tailored rapid cycling synchrotron (RCS) lattice, to overcome the fundamental challenge of extracting multiple beamlets from a single proton bunch for 3D PBS proton FLASH delivery.

The proposed extraction method applies a fast kicker pulse selectively to specific longitudinal segments of the bunch, enabling controlled single-turn extraction.
To physically implement this scheme, a dedicated RCS lattice was designed, with phase advances and optical functions optimized to accommodate the stripline kicker, ESe, and MSe, ensuring technically feasible field strengths and clear transverse separation distance of the extracted beam.
Particle tracking simulations confirm the system's viability.
The results demonstrate that spot dose accuracy can be maintained within clinically acceptable limits under defined conditions, while the inherent beam loss at the septum, arising from the finite rise/fall time of the kicker pulse, is quantified to be around $0.840\%$.
The analysis further identifies longitudinal beam line density as a critical parameter influencing both spot dose accuracy and beam loss at the septum.

In summary, this paper introduced the longitudinal localized kick driven fast extraction method and its corresponding accelerator design, providing a core technical solution for enabling multi-shot, single-bunch extraction in future proton FLASH delivery systems.

\appendix

\begin{widetext}

\section{algorithm used in Sec.~\ref{Sec.2.B} to determine the stripline kicker pulse time}

\begin{algorithm}[H]
	\caption{Kicker Pulse Time Determination Process}
	\label{alg:Kicker_Pulse_Time_Determination_Process}
	\begin{algorithmic}[1]
		\REQUIRE
		\STATE $\tau_{\mathrm{fall}}$: width of the falling edge (\texttt{fall\_time})
		\STATE $\Delta\tau$: width of the time bin (\texttt{bin\_time})
		\STATE $N_{\mathrm{ext}}$: required number of particles to be extracted (\texttt{N\_ext})
		\ENSURE
		\STATE $\tau_1, \tau_2, \tau_3$: flattop start, flattop end (falling edge start), and falling edge end times
		\STATE \texttt{fall\_is\_enough}: boolean flag indicating if falling edge alone provides sufficient particles
		
		\STATE
		\STATE \textbf{Initialize}
		\STATE $\texttt{fall\_is\_enough} \gets \text{false}$
		\STATE $\tau_1, \tau_2, \tau_3 \gets \text{undefined}$
		\STATE $\texttt{fall\_bin\_num} \gets \lceil \tau_{\mathrm{fall}} / \Delta\tau \rceil$
		\STATE Find $\tau_s$ (start bin index) using Eq.~\eqref{eq:Flattop_Begin}
		
		\STATE
		\STATE \textbf{Step 1: Check if falling edge alone provides sufficient particles}
		\FOR{$i = 1$ to $\texttt{fall\_bin\_num}$}
		\STATE Compute $N_{\mathrm{fall}}(i)$ using Eq.~\eqref{eq:N_fall_expression}, integrating from $\tau_s$ to $\tau_s + i \cdot \Delta\tau$
		\IF{$N_{\mathrm{fall}}(i) \geq N_{\mathrm{ext}}$}
		\STATE $\texttt{fall\_is\_enough} \gets \text{true}$
		\STATE $\tau_3 \gets \frac{\Delta\tau \cdot \left(N_{\mathrm{ext}}-N_{\mathrm{fall}}(i-1)\right)}{N_{\mathrm{fall}}(i)-N_{\mathrm{fall}}(i-1)} + \tau_s+(i-1)\cdot\Delta\tau$
		\STATE $\tau_2 \gets \tau_3 - \tau_{\mathrm{fall}}$
		\STATE \textbf{break}
		\ENDIF
		\ENDFOR
		
		\STATE
		\STATE \textbf{Step 2: If falling edge insufficient, include flattop}
		\IF{$\texttt{fall\_is\_enough} = \text{false}$}
		\STATE $N_{\mathrm{flattop}} \gets 0$, $N_{\mathrm{fall}} \gets 0$
		\STATE $N_{\mathrm{total}} \gets N_{\mathrm{flattop}} + N_{\mathrm{fall}}$
		\STATE $j \gets 1$
		\WHILE{$N_{\mathrm{total}} < N_{\mathrm{ext}}$}
		\STATE Compute $N_{\mathrm{flattop}}(j)$ by summing from $\tau_s$ to $\tau_s + j \cdot \Delta\tau$
		\STATE Compute $N_{\mathrm{fall}}(j)$ using Eq.~\eqref{eq:N_fall_expression}, integrating from $\tau_s + j \cdot \Delta\tau$ to $\tau_s + (j + \texttt{fall\_bin\_num}) \cdot \Delta\tau$
		\STATE $N_{\mathrm{total}}(j) \gets N_{\mathrm{flattop}}(j) + N_{\mathrm{fall}}(j)$
		\STATE $j \gets j + 1$
		\ENDWHILE
		\STATE $\tau_2 \gets \frac{\Delta\tau \cdot \left(N_{\mathrm{ext}}-N_{\mathrm{total}}(j-1)\right)}{N_{\mathrm{total}}(j)-N_{\mathrm{total}}(j-1)} + \tau_s+(j-1)\cdot\Delta\tau$
		\STATE $\tau_3 \gets \tau_2 + \tau_{\mathrm{fall}}$
		\STATE $\tau_1 \gets \tau_s$
		\ENDIF
		
		\STATE
		\STATE \textbf{Return} $(\tau_1, \tau_2, \tau_3, \texttt{fall\_is\_enough})$
	\end{algorithmic}
\end{algorithm}
\end{widetext}
\bibliography{ref/apssamp}

@article{Favaudon_2014,
author = {Vincent Favaudon  and Laura Caplier  and Virginie Monceau  and Frédéric Pouzoulet  and Mano Sayarath  and Charles Fouillade  and Marie-France Poupon  and Isabel Brito  and Philippe Hupé  and Jean Bourhis  and Janet Hall  and Jean-Jacques Fontaine  and Marie-Catherine Vozenin },
title = {Ultrahigh dose-rate FLASH irradiation increases the differential response between normal and tumor tissue in mice},
journal = {Science Translational Medicine},
volume = {6},
number = {245},
pages = {245ra93-245ra93},
year = {2014},
doi = {10.1126/scitranslmed.3008973},
URL = {https://www.science.org/doi/abs/10.1126/scitranslmed.3008973}
}

@article{Esplen_2020,
doi = {10.1088/1361-6560/abaa28},
url = {https://doi.org/10.1088/1361-6560/abaa28},
year = {2020},
month = {dec},
publisher = {IOP Publishing},
volume = {65},
number = {23},
pages = {23TR03},
author = {Esplen, Nolan and Mendonca, Marc S and Bazalova-Carter, Magdalena},
title = {Physics and biology of ultrahigh dose-rate (FLASH) radiotherapy: a topical review},
journal = {Physics in Medicine \& Biology}
}

@article{Loo_2024,
  title = {FLASH: New intersection of physics, chemistry, biology, and cancer medicine},
  author = {Vozenin, Marie-Catherine and Loo, Billy W. and Tantawi, Sami and Maxim, Peter G. and Spitz, Douglas R. and Bailat, Claude and Limoli, Charles L.},
  journal = {Rev. Mod. Phys.},
  volume = {96},
  issue = {3},
  pages = {035002},
  numpages = {50},
  year = {2024},
  month = {Sep},
  publisher = {American Physical Society},
  doi = {10.1103/RevModPhys.96.035002},
  url = {https://link.aps.org/doi/10.1103/RevModPhys.96.035002}
}

@article{10.1001/jamaoncol.2022.5843,
    author = {Mascia, Anthony E. and Daugherty, Emily C. and Zhang, Yongbin and Lee, Eunsin and Xiao, Zhiyan and Sertorio, Mathieu and Woo, Jennifer and Backus, Lori R. and McDonald, Julie M. and McCann, Claire and Russell, Kenneth and Levine, Lisa and Sharma, Ricky A. and Khuntia, Dee and Bradley, Jeffrey D. and Simone, Charles B., II and Perentesis, John P. and Breneman, John C.},
    title = {Proton FLASH Radiotherapy for the Treatment of Symptomatic Bone Metastases: The FAST-01 Nonrandomized Trial},
    journal = {JAMA Oncology},
    volume = {9},
    number = {1},
    pages = {62-69},
    year = {2023},
    month = {01},
    issn = {2374-2437},
    doi = {10.1001/jamaoncol.2022.5843},
    url = {https://doi.org/10.1001/jamaoncol.2022.5843}
}

@article{Darafsheh_2025,
doi = {10.1088/1361-6560/add106},
url = {https://doi.org/10.1088/1361-6560/add106},
year = {2025},
month = {may},
publisher = {IOP Publishing},
volume = {70},
number = {10},
pages = {105008},
author = {Darafsheh, Arash and Bey, Anissa},
title = {Implementation of a proton FLASH platform for pre-clinical studies using a gantry-mounted synchrocyclotron},
journal = {Physics in Medicine \& Biology}
}

@article{DIFFENDERFER2020440,
title = {Design, Implementation, and in Vivo Validation of a Novel Proton FLASH Radiation Therapy System},
journal = {International Journal of Radiation Oncology*Biology*Physics},
volume = {106},
number = {2},
pages = {440-448},
year = {2020},
issn = {0360-3016},
doi = {https://doi.org/10.1016/j.ijrobp.2019.10.049},
url = {https://www.sciencedirect.com/science/article/pii/S0360301619340556},
author = {Eric S. Diffenderfer and Ioannis I. Verginadis and Michele M. Kim and Khayrullo Shoniyozov and Anastasia Velalopoulou and Denisa Goia and Mary Putt and Sarah Hagan and Stephen Avery and Kevin Teo and Wei Zou and Alexander Lin and Samuel Swisher-McClure and Cameron Koch and Ann R. Kennedy and Andy Minn and Amit Maity and Theresa M. Busch and Lei Dong and Costas Koumenis and James Metz and Keith A. Cengel}
}

@article{Klimpki_2018,
doi = {10.1088/1361-6560/aacd27},
url = {https://doi.org/10.1088/1361-6560/aacd27},
year = {2018},
month = {jul},
publisher = {IOP Publishing},
volume = {63},
number = {14},
pages = {145006},
author = {Klimpki, G and Zhang, Y and Fattori, G and Psoroulas, S and Weber, D C and Lomax, A and Meer, D},
title = {The impact of pencil beam scanning techniques on the effectiveness and efficiency of rescanning moving targets},
journal = {Physics in Medicine \& Biology}
}

@article{https://doi.org/10.1002/mp.13972,
author = {Kang, M. and Pang, D.},
title = {Commissioning and beam characterization of the first gantry-mounted accelerator pencil beam scanning proton system},
journal = {Medical Physics},
volume = {47},
number = {8},
pages = {3496-3510},
doi = {https://doi.org/10.1002/mp.13972},
url = {https://aapm.onlinelibrary.wiley.com/doi/abs/10.1002/mp.13972},
year = {2020}
}

@article{YOUNKIN2018412,
title = {Multiple energy extraction reduces beam delivery time for a synchrotron-based proton spot-scanning system},
journal = {Advances in Radiation Oncology},
volume = {3},
number = {3},
pages = {412-420},
year = {2018},
issn = {2452-1094},
doi = {https://doi.org/10.1016/j.adro.2018.02.006},
url = {https://www.sciencedirect.com/science/article/pii/S2452109418300319},
author = {James E. Younkin and Martin Bues and Terence T. Sio and Wei Liu and Xiaoning Ding and Sameer R. Keole and Joshua B. Stoker and Jiajian Shen}
}

@article{Roberfroid_2025,
doi = {10.1088/1361-6560/adb9b2},
url = {https://doi.org/10.1088/1361-6560/adb9b2},
year = {2025},
month = {mar},
publisher = {IOP Publishing},
volume = {70},
number = {6},
pages = {065005},
author = {Roberfroid, Benjamin and Chocan Vera, Macarena S and Draguet, Camille and Lee, John A and Barragán-Montero, Ana M and Sterpin, Edmond},
title = {Anticipating potential bottlenecks in adaptive proton FLASH therapy: a ridge filter reuse strategy},
journal = {Physics in Medicine \& Biology}
}

@article{RODDY2024169284,
title = {Design, optimization, and testing of ridge filters for proton FLASH radiotherapy at TRIUMF: The HEDGEHOG},
journal = {Nuclear Instruments and Methods in Physics Research Section A: Accelerators, Spectrometers, Detectors and Associated Equipment},
volume = {1063},
pages = {169284},
year = {2024},
issn = {0168-9002},
doi = {https://doi.org/10.1016/j.nima.2024.169284},
url = {https://www.sciencedirect.com/science/article/pii/S0168900224002109},
author = {David Roddy and Camille Bélanger-Champagne and Sebastian Tattenberg and Stanley Yen and Michael Trinczek and Cornelia Hoehr}
}

@article{Xiong_2025,
doi = {10.1088/1361-6560/ade04a},
url = {https://doi.org/10.1088/1361-6560/ade04a},
year = {2025},
month = {jun},
publisher = {IOP Publishing},
volume = {70},
number = {12},
pages = {125019},
author = {Xiong, Yang and Li, Yan and Yao, Hongjuan and Zheng, Shuxin},
title = {A novel 3D proton pencil beam scanning scheme and key physical design of the corresponding rapid cycling synchrotron for FLASH delivery},
journal = {Physics in Medicine \& Biology}
}

@book{JayFlanz2022,
  author     = {Jay Flanz},
  title      = {Particle Therapy Technology for Safe Treatment},
  address    = {Boca Raton},
  publisher  = {CRC Press},
  year       = {2022},
  pages      = {163},
}

@article{https://doi.org/10.1002/mp.14456,
author = {Folkerts, Michael M. and Abel, Eric and Busold, Simon and Perez, Jessica Rika and Krishnamurthi, Vidhya and Ling, C. Clifton},
title = {A framework for defining FLASH dose rate for pencil beam scanning},
journal = {Medical Physics},
volume = {47},
number = {12},
pages = {6396-6404},
doi = {https://doi.org/10.1002/mp.14456},
url = {https://aapm.onlinelibrary.wiley.com/doi/abs/10.1002/mp.14456},
year = {2020}
}

@article{https://doi.org/10.1002/acm2.12984,
author = {Shang, Charles and Evans, Grant and Rahman, Mushfiqur and Lin, Liyong},
title = {Beam characteristics of the first clinical 360° rotational single gantry room scanning pencil beam proton treatment system and comparisons against a multi-room system},
journal = {Journal of Applied Clinical Medical Physics},
volume = {21},
number = {9},
pages = {266-271},
doi = {https://doi.org/10.1002/acm2.12984},
url = {https://aapm.onlinelibrary.wiley.com/doi/abs/10.1002/acm2.12984},
year = {2020}
}

@article{CHEN2025170431,
title = {Design and construction of the China Spallation Neutron Source},
journal = {Nuclear Instruments and Methods in Physics Research Section A: Accelerators, Spectrometers, Detectors and Associated Equipment},
volume = {1078},
pages = {170431},
year = {2025},
issn = {0168-9002},
doi = {https://doi.org/10.1016/j.nima.2025.170431},
url = {https://www.sciencedirect.com/science/article/pii/S0168900225002323},
author = {Hesheng Chen and Yuanbai Chen and Shinian Fu and Li Ma and Sheng Wang and Fuqing Chen and Yanwei Chen and Haiyi Dong and Lan Dong and Guang Feng and Jun Gu and Lunhua He and Kun He and Wei He and Chunming Hu and Jinshu Huang and Quan Ji and Xuejun Jia and Dapeng Jin and Ling Kang and Wen Kang and Tianjiao Liang and Guopin Lin and Huachang Liu and Jian Li and Huafu Ouyang and Fazhi Qi and Xin Qi and Huamin Qu and Hong Sun and Zhijia Sun and Li Shen and Jingyu Tang and Juzhou Tao and Fangwei Wang and Linshu Wang and Ping Wang and Qingbin Wang and Yaoqing Wu and Jiwei Xi and Taoguang Xu and Wen Yin and Bingyun Zhang and Jing Zhang and Junrong Zhang and Shaoying Zhang and Jingshi Zhao and Luyang Zhao and Yubin Zhao and Min Zhou and Tao Zhu and Jian Zhuang}
}

@article{https://doi.org/10.1118/1.3187229,
author = {Smith, Alfred and Gillin, Michael and Bues, Martin and Zhu, X. Ronald and Suzuki, Kazumichi and Mohan, Radhe and Woo, Shiao and Lee, Andrew and Komaki, Ritsko and Cox, James and Hiramoto, Kazuo and Akiyama, Hiroshi and Ishida, Takayuki and Sasaki, Toshie and Matsuda, Koji},
title = {The M. D. Anderson proton therapy system},
journal = {Medical Physics},
volume = {36},
number = {9Part1},
pages = {4068-4083},
doi = {https://doi.org/10.1118/1.3187229},
url = {https://aapm.onlinelibrary.wiley.com/doi/abs/10.1118/1.3187229},
year = {2009}
}

@article{COMBS201041,
title = {Particle therapy at the Heidelberg Ion Therapy Center (HIT) – Integrated research-driven university-hospital-based radiation oncology service in Heidelberg, Germany},
journal = {Radiotherapy and Oncology},
volume = {95},
number = {1},
pages = {41-44},
year = {2010},
issn = {0167-8140},
doi = {https://doi.org/10.1016/j.radonc.2010.02.016},
url = {https://www.sciencedirect.com/science/article/pii/S016781401000109X},
author = {Stephanie E. Combs and Oliver Jäkel and Thomas Haberer and Jürgen Debus}
}

@ARTICLE{Cardona1221935,
  author={Cardona, J. and Peggs, S. and Kewisch, J.},
  journal={IEEE Transactions on Nuclear Science}, 
  title={Optical design of the rapid cycling medical synchrotron}, 
  year={2003},
  volume={50},
  number={4},
  pages={1147-1152},
  doi={10.1109/TNS.2003.815103}}

@INPROCEEDINGS{Yamaguchi753429,
  author={Yamaguchi, A. and Nakayama, K. and Rizawa, T. and Sukenobu, S. and Satoh, K. and Morii, Y. and Tanabe, Y. and Chiba, Y.},
  booktitle={Proceedings of the 1997 Particle Accelerator Conference (Cat. No.97CH36167)}, 
  title={A compact proton accelerator system for cancer therapy}, 
  year={1997},
  volume={3},
  number={},
  pages={3828-3830 vol.3},
  doi={10.1109/PAC.1997.753429}}

@article{Fraser:2018bcq,
    author = {Fraser, M. A.},
    editor = {Holzer, Bernhard},
    title = {Fast extraction: single and multi-turn},
    doi = {10.23730/CYRSP-2018-005.285},
    journal = {CERN Yellow Rep. School Proc.},
    volume = {5},
    pages = {285},
    year = {2018}
}

@article{PhysRevAccelBeams.28.020401,
  title = {Design of the EIC hadron storage ring stripline injection kicker with a novel impedance tuning capability},
  author = {Sangroula, M. P. and Liaw, C. J. and Liu, C. and Sandberg, J. and Tsoupas, N. and Xiao, B. and Sun, X.},
  journal = {Phys. Rev. Accel. Beams},
  volume = {28},
  issue = {2},
  pages = {020401},
  numpages = {13},
  year = {2025},
  month = {Feb},
  publisher = {American Physical Society},
  doi = {10.1103/PhysRevAccelBeams.28.020401},
  url = {https://link.aps.org/doi/10.1103/PhysRevAccelBeams.28.020401}
}

@article{WU201845,
title = {Development of the CSNS Lambertson magnet with very low stray field},
journal = {Nuclear Instruments and Methods in Physics Research Section A: Accelerators, Spectrometers, Detectors and Associated Equipment},
volume = {882},
pages = {45-52},
year = {2018},
issn = {0168-9002},
doi = {https://doi.org/10.1016/j.nima.2017.10.093},
url = {https://www.sciencedirect.com/science/article/pii/S0168900217311828},
author = {Yuwen Wu and Wen Kang and Yuan Chen and Xi Wu and Shuai Li and Lei Wang and Changdong Deng and Li Li and Jianxin Zhou and Yiqin Liu}
}

@proceedings{Brandt:1071486,
      author        = "Brandt, Daniel",
      title         = "{CAS - CERN Accelerator School: Beam Diagnostics}",
      organization  = "CERN",
      publisher     = "CERN",
      address       = "Geneva",
      year          = "2009",
      url           = "https://cds.cern.ch/record/1071486",
      note          = "In collaboration with Synchrotron SOLEIL, Gif-sur-Yvette,
                       France",
      doi           = "10.5170/CERN-2009-005",
}

@book{Badano_385378,
      author = {Badano, L and Benedikt, Michael and Bryant, P J and Crescenti, M and Holy, P and Maier, A T and Pullia, M and Rossi, S and Knaus, P},
      collaboration = {CERN-TERAFoundation-MedAustronOncology-2000},
      title = {Proton-Ion Medical Machine Study (PIMMS), 1},
      year = {2000},
      url = {https://cds.cern.ch/record/385378}
}

@ARTICLE{6068232,
  author={Kang, W. and Deng, C. D. and Li, Q. and Li, L. and Chen, W. and Yin, B. G. and Zhou, J. X. and Sun, X. J. and Chen, F. S. and Shi, C. T.},
  journal={IEEE Transactions on Applied Superconductivity}, 
  title={Research and Development of the AC Magnets for CSNS/RCS}, 
  year={2012},
  volume={22},
  number={3},
  pages={4001204-4001204},
  doi={10.1109/TASC.2011.2174543}
}

@InProceedings{Yao:IPAC2019-WEPTS033,
  author       = {H.J. Yao and others},
  title        = {A High-performance Code for Beam Dynamics Simulation of Synchrotrons},
  booktitle    = {Proc. 10th International Particle Accelerator Conference (IPAC'19),
                  Melbourne, Australia, 19-24 May 2019},
  pages        = {3170--3173},
  paper        = {WEPTS033},
  language     = {english},
  venue        = {Melbourne, Australia},
  series       = {International Particle Accelerator Conference},
  number       = {10},
  publisher    = {JACoW Publishing},
  address      = {Geneva, Switzerland},
  month        = {Jun.},
  year         = {2019},
  isbn         = {978-3-95450-208-0},
  doi          = {10.18429/JACoW-IPAC2019-WEPTS033},
  url          = {http://jacow.org/ipac2019/papers/wepts033.pdf},
  note         = {https://doi.org/10.18429/JACoW-IPAC2019-WEPTS033},
}

@article{LIU2020163670,
title = {Development of a stripline kicker for the Hefei Advanced Light Facility},
journal = {Nuclear Instruments and Methods in Physics Research Section A: Accelerators, Spectrometers, Detectors and Associated Equipment},
volume = {961},
pages = {163670},
year = {2020},
issn = {0168-9002},
doi = {10.1016/j.nima.2020.163670},
url = {https://www.sciencedirect.com/science/article/pii/S0168900220302382},
author = {Wei Liu and Lei Shang and Wenbin Song and Fenglei Shang and Zhenbiao Sun}
}

@article{https://doi.org/10.1002/mp.13622,
author = {Arjomandy, Bijan and Taylor, Paige and Ainsley, Christopher and Safai, Sairos and Sahoo, Narayan and Pankuch, Mark and Farr, Jonathan B. and Yong Park, Sung and Klein, Eric and Flanz, Jacob and Yorke, Ellen D. and Followill, David and Kase, Yuki},
title = {AAPM task group 224: Comprehensive proton therapy machine quality assurance},
journal = {Medical Physics},
volume = {46},
number = {8},
pages = {e678-e705},
doi = {https://doi.org/10.1002/mp.13622},
url = {https://aapm.onlinelibrary.wiley.com/doi/abs/10.1002/mp.13622},
year = {2019}
}

@InProceedings{Sterbini:IPAC2014-THPME173,
  author       = {G. Sterbini and B. Dehning and S.S. Gilardoni and A. Guerrero},
  title        = {{B}eam{-}based {M}easurements of the {CPS} {W}ire {S}canner {P}recision and {A}ccuracy},
  booktitle    = {Proc. 5th International Particle Accelerator Conference (IPAC'14),
                  Dresden, Germany, June 15-20, 2014},
  pages        = {3674--3676},
  paper        = {THPME173},
  language     = {english},
  venue        = {Dresden, Germany},
  series       = {International Particle Accelerator Conference},
  number       = {5},
  publisher    = {JACoW},
  address      = {Geneva, Switzerland},
  month        = {July},
  year         = {2014},
  isbn         = {978-3-95450-132-8},
  doi          = {https://doi.org/10.18429/JACoW-IPAC2014-THPME173},
  url          = {http://jacow.org/ipac2014/papers/thpme173.pdf},
  note         = {https://doi.org/10.18429/JACoW-IPAC2014-THPME173},
}

@InProceedings{Yao:NAPAC2016-WEPOB24,
  author       = {C. Yao and others},
  title        = {{P}reliminary {T}est {R}esults of a {P}rototype {F}ast {K}icker for {APS} {MBA} {U}pgrade},
  booktitle    = {Proc. of North American Particle Accelerator Conference (NAPAC'16),
                  Chicago, IL, USA, October 9-14, 2016},
  pages        = {950--952},
  paper        = {WEPOB24},
  language     = {english},
  venue        = {Chicago, IL, USA},
  series       = {North American Particle Accelerator Conference},
  number       = {3},
  publisher    = {JACoW},
  address      = {Geneva, Switzerland},
  month        = {Jan.},
  year         = {2017},
  isbn         = {978-3-95450-180-9},
  doi          = {https://doi.org/10.18429/JACoW-NAPAC2016-WEPOB24},
  url          = {https://jacow.org/napac2016/papers/wepob24.pdf},
  note         = {https://doi.org/10.18429/JACoW-NAPAC2016-WEPOB24},
}

@InProceedings{PASJ2016MOP110,
  author       = {Chikaori Mitsuda and Teruo Honiden and Kazuo Kobayashi and Toshiaki Kobayashi and Shigeki Sasaki and Norio Sekine},
  title        = {Development of SiC solid-state fast pulse driver using stripline type Blumlein in SPring-8},
  booktitle    = {Proc. 13th Annual Meeting of Particle Accelerator Society of Japan,
                  August 8-10, 2016, Chiba, Japan},
  pages        = {694--698},
  paper        = {MOP110},
  series       = {Annual Meeting of Particle Accelerator Society of Japan},
  address      = {Chiba, Japan},
  month        = {Aug.},
  year         = {2016}
}

@article{PhysRevSTAB.14.051002,
  title = {Multibunch beam extraction using the strip-line kicker at the KEK Accelerator Test Facility},
  author = {Naito, T. and Araki, S. and Hayano, H. and Kubo, K. and Kuroda, S. and Terunuma, N. and Okugi, T. and Urakawa, J.},
  journal = {Phys. Rev. ST Accel. Beams},
  volume = {14},
  issue = {5},
  pages = {051002},
  numpages = {9},
  year = {2011},
  month = {May},
  publisher = {American Physical Society},
  doi = {10.1103/PhysRevSTAB.14.051002},
  url = {https://link.aps.org/doi/10.1103/PhysRevSTAB.14.051002}
}

@article{WOS:000813996500001,
Author = {Liu, Hanyang and Wang, Sheng},
Title = {Longitudinal beam dynamic design of 500 kW beam power upgrade for
   CSNS-II RCS},
Journal = {RADIATION DETECTION TECHNOLOGY AND METHODS},
Year = {2022},
Volume = {6},
Number = {3},
Pages = {339-348},
Month = {SEP},
DOI = {10.1007/s41605-022-00325-5},
EarlyAccessDate = {JUN 2022},
ISSN = {2509-9930},
EISSN = {2509-9949},
Unique-ID = {WOS:000813996500001},
}

@article{BAUDRENGHIEN2013181,
title = {Longitudinal emittance blowup in the large hadron collider},
journal = {Nuclear Instruments and Methods in Physics Research Section A: Accelerators, Spectrometers, Detectors and Associated Equipment},
volume = {726},
pages = {181-190},
year = {2013},
issn = {0168-9002},
doi = {https://doi.org/10.1016/j.nima.2013.05.060},
url = {https://www.sciencedirect.com/science/article/pii/S0168900213006669},
author = {P. Baudrenghien and T. Mastoridis}
}

@article{Prezado_2024,
doi = {10.1088/1361-6560/ad4192},
url = {https://doi.org/10.1088/1361-6560/ad4192},
year = {2024},
month = {may},
publisher = {IOP Publishing},
volume = {69},
number = {10},
pages = {10TR02},
author = {Prezado, Yolanda and Grams, Michael and Jouglar, Emmanuel and Martínez-Rovira, Immaculada and Ortiz, Ramon and Seco, Joao and Chang, Sha},
title = {Spatially fractionated radiation therapy: a critical review on current status of clinical and preclinical studies and knowledge gaps},
journal = {Physics in Medicine \& Biology}
}

@article{LI2024737,
title = {Overview and Recommendations for Prospective Multi-institutional Spatially Fractionated Radiation Therapy Clinical Trials},
journal = {International Journal of Radiation Oncology*Biology*Physics},
volume = {119},
number = {3},
pages = {737-749},
year = {2024},
issn = {0360-3016},
doi = {https://doi.org/10.1016/j.ijrobp.2023.12.013},
url = {https://www.sciencedirect.com/science/article/pii/S0360301623082469},
author = {Heng Li and Nina A. Mayr and Robert J. Griffin and Hualin Zhang and Damodar Pokhrel and Michael Grams and Jose Penagaricano and Sha Chang and Matthew B. Spraker and James Kavanaugh and Liyong Lin and Khadija Sheikh and Sina Mossahebi and Charles B. Simone and David Roberge and James W. Snider and Pouya Sabouri and Andrea Molineu and Ying Xiao and Stanley H. Benedict}
}

\end{document}